\documentclass[sigconf]{acmart} 

\usepackage{subcaption} 
\usepackage{color}
\usepackage{verbatim} 
\usepackage{multirow}
\usepackage{algorithm}
\usepackage[noend]{algorithmic}
\usepackage{xspace}
\usepackage{graphicx}
\usepackage{epsfig}
\usepackage{amsmath}
\usepackage{amssymb}
\usepackage{float}
\usepackage{enumitem}
\usepackage{balance}
\usepackage[font=bf,belowskip=0pt,aboveskip=5pt]{caption}
\usepackage{lipsum}
\usepackage{bm}
\PassOptionsToPackage{hyphens}{url}
\usepackage{hyperref}
\usepackage{url}
\hypersetup{colorlinks,citecolor=blue}
\usepackage{times}
\usepackage[T1]{fontenc}

\usepackage{amsmath}
\makeatletter
\g@addto@macro\normalsize{%
  \setlength\abovedisplayskip{2pt}
  \setlength\belowdisplayskip{2pt}
  \setlength\abovedisplayshortskip{2pt}
  \setlength\belowdisplayshortskip{2pt}
}

\usepackage{epigraph} 

\usepackage{etoolbox}
\apptocmd{\thebibliography}{\normalsize}{}{}

\usepackage{booktabs} 

\newcommand{\hide}[1]{}
\newcommand{\xhdr}[1]{\vspace{1.0mm}\noindent{{\bf #1.}}}
\newcommand{\xhdrnodot}[1]{\vspace{1.0mm}\noindent{{\bf #1}}}

\newcommand{\eg}{\emph{e.g.}}
\newcommand{\ie}{\emph{i.e.}}

\setitemize{noitemsep,topsep=0pt,parsep=0pt,partopsep=0pt,leftmargin=*}
\setenumerate{noitemsep,topsep=0pt,parsep=0pt,partopsep=0pt,leftmargin=*}

\graphicspath{{./FIG/}}

\begin{document}

\copyrightyear{2018}
\acmYear{2018} 
\setcopyright{iw3c2w3}
\acmConference[WWW 2018]{The 2018 Web Conference}{April 23--27, 2018}{Lyon, France}
\acmBooktitle{WWW 2018: The 2018 Web Conference, April 23--27, 2018, Lyon, France}
\acmPrice{}
\acmDOI{10.1145/3178876.3186062}
\acmISBN{978-1-4503-5639-8/18/04.}

\fancyhead{}

\title{I'll Be Back: On the Multiple Lives of Users\\of a Mobile Activity Tracking Application}
\author{Zhiyuan Lin}
\affiliation{%
  \institution{Stanford University}
}
\email{zylin@cs.stanford.edu}

\author{Tim Althoff}
\affiliation{%
  \institution{Stanford University}
}
\email{althoff@cs.stanford.edu}

\author{Jure Leskovec}
\affiliation{%
  \institution{Stanford University}
}
\email{jure@cs.stanford.edu}

\renewcommand{\shorttitle}{The Multiple Lives of Users of a Mobile Activity Tracking Application}

\begin{abstract}


Mobile health applications that track activities, such as exercise, sleep, and diet, are becoming widely used.
While these activity tracking applications have the potential to improve our health,
user engagement and retention are critical factors for their success.
However, long-term user engagement patterns in real-world activity tracking applications are not yet well understood.
%


Here we study 
user engagement patterns
within a mobile physical activity tracking application consisting of 115 million logged activities taken by over a million users over 31 months.
Specifically, we show that over 75\% of users return and re-engage with the application after prolonged periods of inactivity, no matter the duration of the inactivity.

  We find a surprising result that the re-engagement usage patterns resemble those of the \textit{start} of the initial engagement period, rather than being a simple continuation of the \textit{end} of the initial engagement period.
This evidence points to a conceptual model of \emph{multiple lives} of user engagement, extending the prevalent \emph{single life} view of user activity.
We demonstrate that these multiple lives occur because the users have a variety of different \emph{primary intents or goals} for using the app.
These primary intents are associated with how long each life lasts and how likely the user is to re-engage for a new life. 
We find evidence for users being more likely to stop using the app once they achieved their primary intent or goal (\eg, weight loss). However, these users might return once their original intent resurfaces (\eg, wanting to lose newly gained weight).
We discuss implications of the \emph{multiple life paradigm} and propose a novel prediction task of predicting the number of lives of a user.
Based on insights developed in this work, including a marker of improved primary intent performance, our prediction models achieve 71\% ROC AUC.
Overall, our research has implications for modeling user re-engagement in health activity tracking applications and has consequences for how notifications, recommendations as well as gamification can be used to increase engagement.

\end{abstract}

\maketitle




\section{Introduction}
\label{sec:intro}

Activity tracking applications for mobile health have become an important part of people's daily lives.
A US-nationwide study in 2013 found that 69\% of US adults keep track of one or more health indicators including weight, diet, exercise, or symptoms, and 21\% among them used a mobile app or device to do so~\cite{fox2013tracking}.
Globally, the mobile health market is projected market to grow to 
over \$500 billion by 2025~\cite{tmr2017digitalhealthmarket}.
Understanding how users engage with mobile activity tracking applications has the potential to signficantly improve 
people's health, for instance by preventing negative health outcomes and promoting the adoption and maintenance of healthy behaviors~\cite{nahum2016just,thomas2015behavioral,althoff2017population}.

However, user engagement patterns in these activity tracking applications, especially long-term and at scale, are not yet well understood.
While there has been a wealth of research on user engagement in various \textit{online} settings (\eg,~\cite{lalmas2014measuring,attfield2011towards,obrien2008user,obrien2011exploring,saveski2014geography, shafiq2014understanding}),
it is unclear how users engage and re-engage with activity tracking applications that capture their \textit{offline} lives in the real world. 
Understanding user engagement in these real-world contexts is particularly important given that we could help people improve their health.
As mobile health applications that track a variety of health metrics and daily activities are becoming more popular, it is important to model and understand user engagement in these applications.





Typical modeling of user engagement considers users to have a single ``lifetime'' (or ``lifespan'') during which the user typically becomes less and less engaged on the platform~\cite{yang2010activity,danescu2013no,anderson2014dynamics,benson2016modeling,trouleau2016just,kapoor2015just}.
This conceptual model of a \emph{single lifetime} has been widely used in user modeling and engagement research, as well as interventions aimed at increasing user engagement~\cite{althoff2017onlineactions,anderson2014engaging,anderson2013steering,anderson2014dynamics,benson2016modeling,trouleau2016just,kapoor2015just}.
For example, existing work has focused on predicting the duration of user's single lifetime~\cite{kapoor2014hazard,jing2017neural,danescu2013no,wang2016accurate} and attempted to extend a user's lifetime to \emph{retain} them~\cite{althoff2015donor,yang2010activity}.
Usually, once a user has been absent for a long time, they are very unlikely to return and thus pronounced ``dead''~\cite{leskovec2008microscopic}.
Survival modeling techniques, which have been applied to user modeling~\cite{dave2017fast,kapoor2014hazard,jing2017neural}, also assume that once a user has ``died'', they are not coming back to re-engage with the app.
%
However, in the context of activity tracking applications that track real-world behaviors, user engagement patterns may not follow this single lifetime model.
Users may return to re-engage with applications after long periods of inactivity and, this way, start a new life.  
There is a limited understanding of how users re-engage with activity tracking apps after long periods of inactivity, the mechanisms behind such behavior, and whether this behavior may be predictable ahead of time.

\xhdr{This work}
Here, we conduct a large-scale observational study of user re-engagement patterns within a mobile activity tracking application. 
We demonstrate empirically, across 115 million logged activities taken by over a million users over 31 months, that over 75\% of active users do return to the application after a prolonged period of inactivity. 
This behavior is independent of the duration of inactivity and often users return to the application multiple times. 
While many applications use notifications and e-mails to regain user attention after brief periods of inactivity (\eg, a few days), we observe a large fraction of users returning after much longer periods of time (\eg, 90 days). 
Importantly, the usage patterns when re-engaging with the app mimic those of the \emph{start} of the user's initial engagement period and cannot be explained as a simple continuation of the end of the initial engagement period. 
These observations suggest a conceptual model of user engagement exhibiting \emph{multiple lives},
 where user activity after re-engagement is more similar to the beginning of the previous life than the end of it.
Users re-engage with this application after prolonged inactivity to begin a \emph{new life}, rather than continuing patterns of their previous life---often again and again.
This multiple life paradigm stands in contrast to the single life paradigm prevalent in existing work on user lifetime modeling.

Despite a variety of available activities in the app, we find that users generally focus on logging a single primary activity, reflecting the user's goal or \emph{primary intent}. 
We show that multiple lives occur because users have a variety of different primary intents for using the mobile health app.
While primary intents are different from user to user, the intents stay relatively constant over multiple lives of the same user. 
Primary intents shape how long each life lasts and how likely the user is to return for another new life.
Associated with the user's intent are performance goals such as losing body weight, walking or running more, or sleeping longer. 
We demonstrate that users getting closer or reaching such goals are more likely to stop using the app; that is, the app has fulfilled its purpose for them.
However, users regularly return after prolonged inactivity with a lower level of performance (\eg, body weight increased again), perhaps after a setback or after setting a new goal; that is, the app may become useful again to pursue an old or new goal. 
We further demonstrate that user re-engagement may be driven by external and seasonal factors such as New Year's resolutions in January, and summer months during which users may be more interested in physical activity and losing weight. 
Multiple life patterns also vary with user demographics. 
For example, we find that young users have much shorter lifetimes and are less likely to return for another life than older users.

To demonstrate the predictive power of our observations, we also formulate a novel prediction task of predicting how many lives a user will have. We demonstrate that the insights developed in this work allow us to predict whether a user re-engages for another life with 71\% ROC AUC.
We also show that improving performance related to the user's primary intent is a marker of likely leaving the app soon, because their primary goal may have been fulfilled.

While we demonstrate the \emph{multiple lives paradigm} within a single mobile activity tracking application, our conceptual model may be applicable to other mobile health applications, in particular when users may be pursuing, reaching, and resetting goals.
Overall, our findings around the multiple lives paradigm have important implications for increasing user engagement in activity tracking applications.
For example, early recognition of the primary intent could enable a more engaging personalized user experience and better notification or gamification experience.
Further, recognizing that certain usage intents lead to few and short lifetimes can highlight which experiences in the app are currently not well supported.
In addition, intents associated with many but very short lives could be a prime target to win users back.

\section{Existence of Multiple Lives}
\label{sec:multiple_lives}

In this section, we describe our dataset from an activity tracking application and demonstrate that user engagement with the application is segmented into multiple disjoint active periods with distinct characteristics. 


\begin{table}[tbp]
\centering
\resizebox{.85\columnwidth}{!}{%
\begin{tabular}{lrr}
  \toprule
  Dataset Statistics  \\ \midrule
  Observation period & 31 months  \\
                 & July 2013 - Jan 2016 \\
  \# total users & 1,329,767 \\
  \# total check-ins & 114,947,892 \\
  Avg. \# of active periods ($\delta$ = 30 days) & 1.7  \\
  Avg. inactive period duration ($\delta$ = 30 days) & 102.3 days\\
  Avg. active period duration ($\delta$ = 30 days) & 22.4 days\\
  Median age & 32 years\\
  \% users female & 48.5\% \\
  \% underweight (BMI $<$ 18.5) & 4.2\% \\
  \% normal weight (18.5 $\leq$ BMI $<$ 25) & 43.6\% \\
  \% overweight (25 $\leq$ BMI $<$ 30) & 31.0\% \\
  \% obese (30 $\leq$ BMI) & 21.3\% \\
  \bottomrule
 \end{tabular}
 }
 \caption{Dataset statistics.}
 \label{tab:dataset}
  \vspace{-4mm}
\end{table}

\subsection{Dataset}
We conduct an observational study using data from a mobile activity tracking application, Argus by Azumio~\cite{althoff2017onlineactions,althoff2017large,shameli2017gamification} (Table~\ref{tab:dataset}). 
This smartphone app allows users to track various daily activities including
running, walking, cardio, heart rate, weight, sleep, drink, and food logging activities (the app also supports other more rarely used activities such as measuring stress or logging yoga which we aggregate and call ``other'' activities; less than 2\% of total logging).
For example, the drink activity is used to keep track of the user's daily fluid intake and the workout activity is used to log various indoor exercises such as weightlifting or indoor-cycling.
In addition, the app passively logs steps through the phone's accelerometers and automatically infers calories consumption for users.
However, we do not consider passively logged activity 
as active engagement and thus filter out such data. 
We call the user action of logging a particular activity at a particular time a {\em check-in}. 
We focus on users who use the app for at least a week. 
Our final dataset~\footnote{All data analyzed is pre-existing and de-identified. We have also obtained necessary IRB approvals.} includes over one million users actively logging 115 million check-ins over the course of 31 months (Table~\ref{tab:dataset}).
This long observation period and large scale uniquely enables us to study re-engagement patterns after prolonged user absence.
Due to the popularity of the app, its users are relatively diverse in terms of age, gender, weight status, country of origin, and other features~\cite{althoff2017large}.


\begin{figure}[tb]
\vspace{-5.5mm}
\begin{subfigure}{.45\textwidth}
  \includegraphics[width=\textwidth]{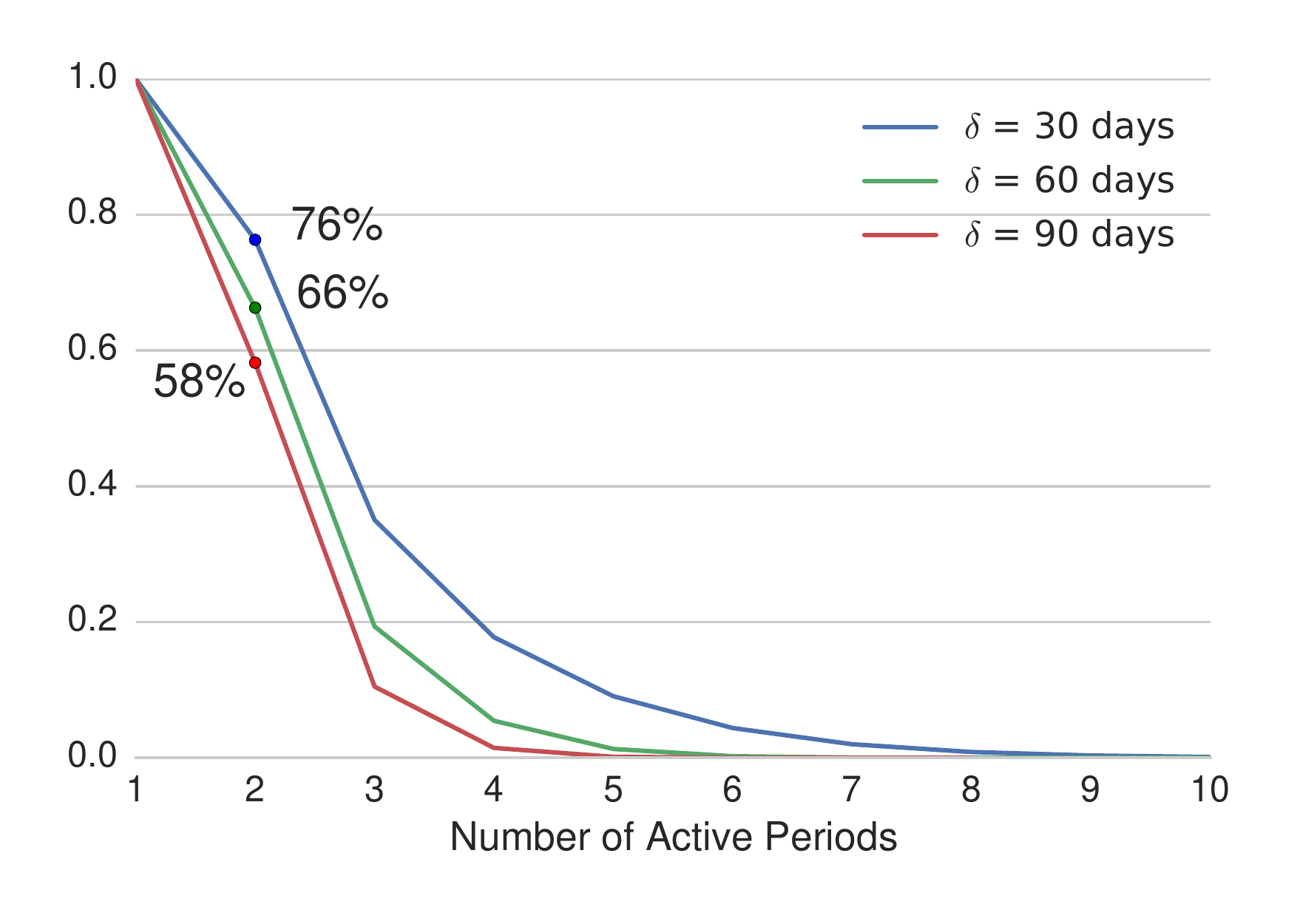}
  \vspace{-8mm}
  \caption{CCDF of number of active periods per user.}
     \label{fig:ccdf_number_of_lives_dist}
\end{subfigure}
\hfill
\begin{subfigure}{.24\textwidth}
  \centering
  \includegraphics[width=\textwidth]{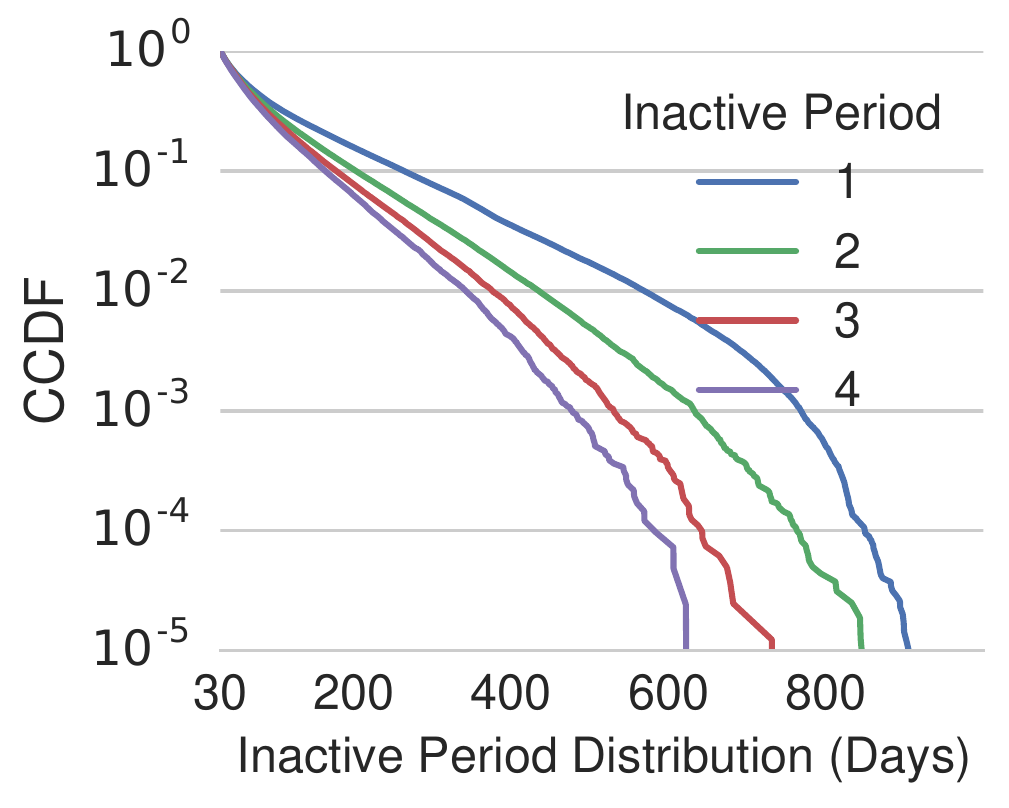}
  \caption{Duration of inactive periods.}
     \label{fig:inter_life_gap}
\end{subfigure}
\begin{subfigure}{.23\textwidth}
  \includegraphics[width=\textwidth]{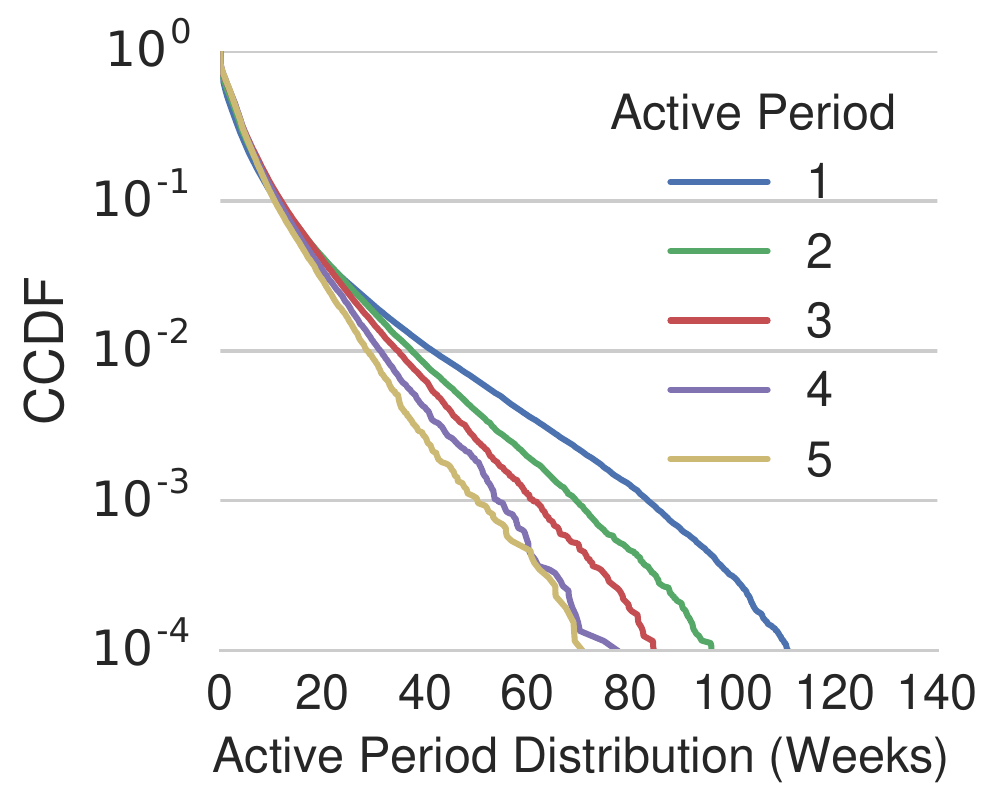}
  \caption{Duration of active periods.}
     \label{fig:lifetime_dist}
\end{subfigure}

\caption{Users have multiple active periods. 
(a) 76\% of users re-engage with the app for a second active period or more (after being absent for more than $\bf{\delta=30}$ days). 
(b) The gap between consecutive active periods (inactive period) becomes shorter over time.
(c) Active periods also become shorter over time.
CCDF refers to Complementary Cumulative Distribution Function.
} 
\label{fig:multiple_lives_dist}
\vspace{-3mm}
\end{figure}

\subsection{Multiple lives}
\label{subsec:multiple_lives}
First, we demonstrate that user engagement with the application is segmented into multiple disjoint \emph{active periods}.
We define an active period as a maximal segment with at most $\delta$ days in between consecutive user's check-ins. 
If a user has not logged an action for at least $\delta$ days, we refer to it as an \emph{inactive period}.
An inactive period between two consecutive active periods is longer than $\delta$ days by definition.
We use ``(in-)activity index'' to refer to a user's $n$-th (in-)active period.
We calculate the duration of an active period as the number of days between the first and last check-in in the period. 
We consider only users with $\lambda$ days of total lifetime, where $\lambda > \delta$, such that it would be at least theoretically possible for a user to have more than one active period at a given threshold~$\delta$. Otherwise, we would, by design, obtain a downwardly biased estimate of the number of active periods of a user.

\xhdr{Results}
We find that over 75\% of users (total lifetime $\lambda$>30 days) become active again after a prolonged $\delta$=30 days of inactivity (Figure~\ref{fig:ccdf_number_of_lives_dist}). 
And these users often return more than once to the app. 
For example, 59\% of users have at least three active periods where each time they were absent from the app for more $\delta$=30 days. 
Importantly, this dynamics does not depend on the definition of the inactivity gap $\delta$.
Even when users are gone for $\delta$=90 days, we find that 58\% of them return for at least one more active period.
Thus, even for very large inactivity thresholds $\delta$ users have \textit{multiple} active periods, 
and this not simply an artifact of a small value of $\delta$. 
We emphasize that, no matter the particular inactivity threshold, \emph{most} users return to have multiple active periods. 
Therefore, studying multiple active periods does not restrict us to a small subsample of the population but, in fact, is relevant to the vast majority of users.
We note that many applications use notifications and e-mails to regain user attention after brief periods of inactivity (\eg, a few days). 
However, we observe a large fraction of users returning after much longer periods of time (\eg, 90 days), when these users would typically have been considered ``dead'' for all practical purposes.
In the following, we use $\delta$=30 days unless specified otherwise.

We find that duration of both active as well as inactive period decrease with each additional active period as shown in Figures~\ref{fig:inter_life_gap} and \ref{fig:lifetime_dist}, respectively.
The average  duration of the \emph{first} active period is 24 days and the average inactive period duration between the first and second active period is 114 days long. 
Note the log scale of the $y$-axis and the partially linear behavior of all curves. 
This suggests that both inactive as well as active period duration distributions can be approximated by exponential distributions.

\begin{figure}[tb]
\vspace{-2mm}
\begin{subfigure}{.23\textwidth}
  \includegraphics[width=\textwidth]{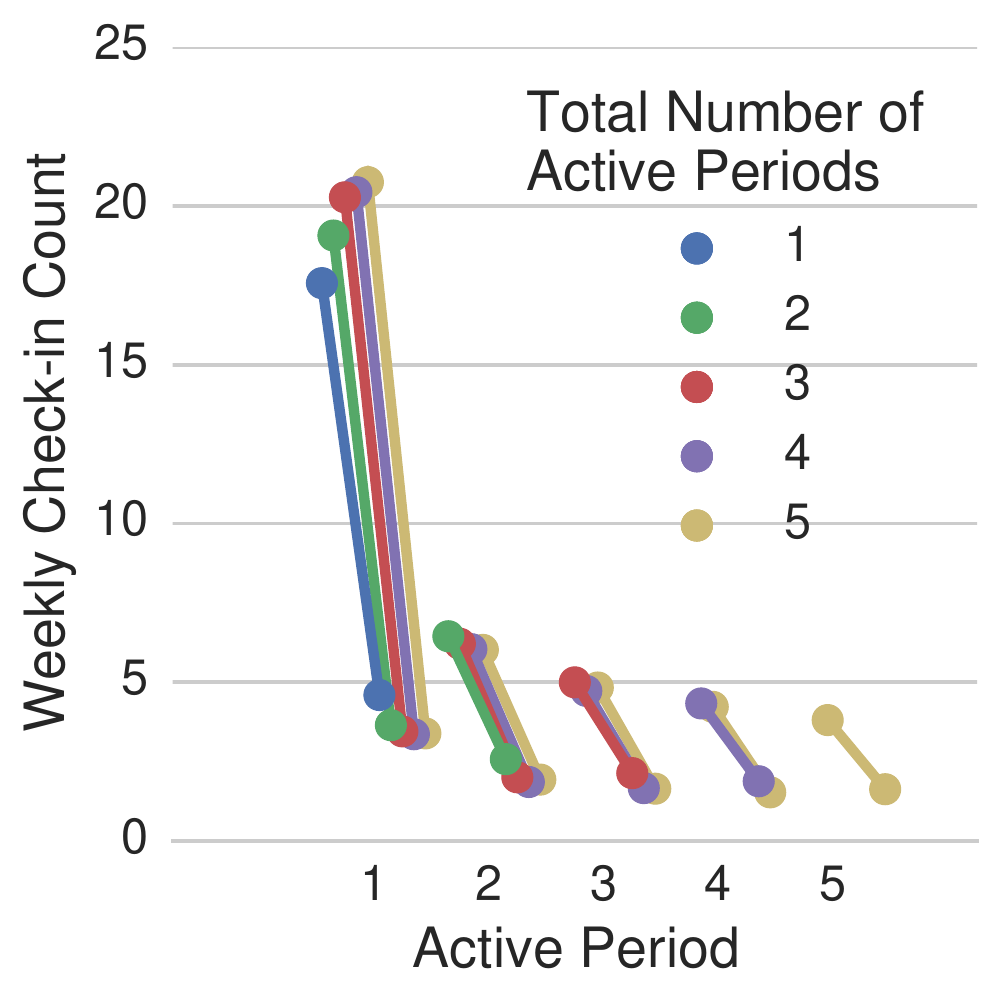}
  \vspace{-6mm}
  \caption{First and last full week's check-in count.}
     \label{fig:zig_zag_checkin_count}
\end{subfigure}
\hfill
\begin{subfigure}{.23\textwidth}
  \includegraphics[width=\textwidth]{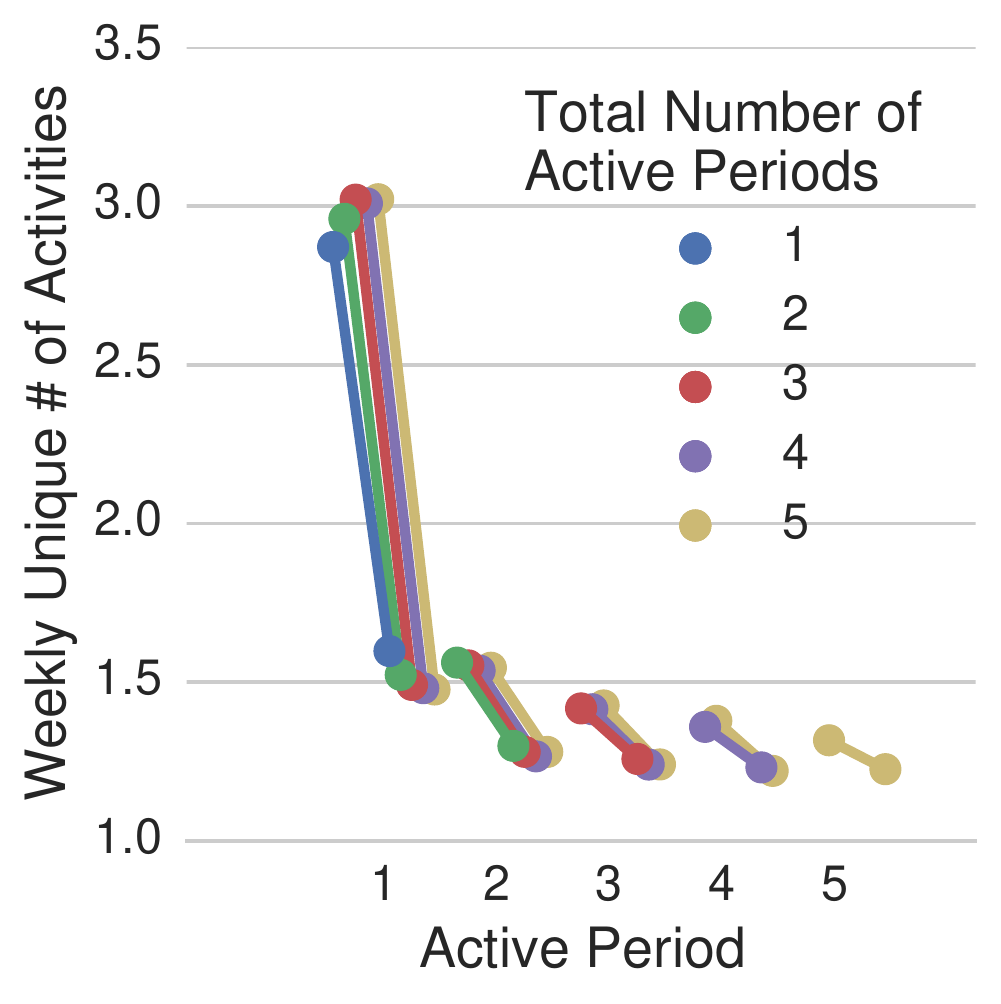}
  \caption{Number of Unique Activities}
     \label{fig:zig_zag_unique_activity}
\end{subfigure}
\caption{
Users start active periods with frequent and diverse check-ins but end them with fewer and less diverse check-ins.
When they re-engage with the app in another active period, check-ins are again more frequent and diverse again.
Error bars in all plots correspond to bootstrapped 95\% confidence intervals. Here they are too small to be visible.} 
\label{fig:zig_zag_summary_user_behavior}
\vspace{-3mm}
\end{figure}

\xhdrnodot{Are multiple active periods simply segments of a single, long, fragmented life?}
We consider two simple measures of user activity: The number of check-ins per week and the number of distinct activities per week as a measure of diversity. 
(We find similar results when using entropy of activity distribution or fraction of check-ins of most frequent activity instead.) 
In Figure~\ref{fig:zig_zag_checkin_count}, we plot the number of check-ins per week for every active period's first and last week, connected by a line.active period
Figure~\ref{fig:zig_zag_unique_activity} shows the same for the number of distinct activities per week. 

In both plots we observe a ``zig-zag'' pattern demonstrating that users start each active period with a larger number of check-ins compared to when they ended their previous active period, introducing a behavior discontinuity. 
Furthermore, these check-ins are also more diverse. 
Users also end each active period being less engaged and focusing on a less diverse set of activities than at the start of their active period.
Importantly, note that the usage patterns at the \emph{start} of a new active period mimic those of the \emph{start} of the previous active period and are not simply a continuation of the \emph{end} of the previous active periods.
The observation suggests that a user's active periods should in fact be considered separately, instead of considering them to be a single, long, fragmented life of a user.
We observe this pattern for a wide range of potential $\delta$ values, which further supports this claim.

Our findings point to a very interesting pattern of user engagement with a health activity tracking app. We observe that most users' lifetimes can be segmented into multiple active periods (average duration of 24 days) separated my long periods of inactivity (average duration 114 days). Furthermore, we observe that users often return after a long absence and that their activity is not a continuation of the previous usage patterns but it looks like as if they are using the app for the first time (Figure~\ref{fig:zig_zag_summary_user_behavior}).

Conceptually, this is consistent with a \textit{multiple lives} metaphor, where user's leave the app, are absent for a long time, and then return as if they are a new user.
Thus, we will use ``life'' to refer to an individual active period, 
use ``lifetime'' to refer to the duration of an active period, 
and use ``life index'' to refer to a user's n-th active period.
In the rest of the paper, we show how this multiple lives paradigm affects health tracking app usage\footnote{Of course, one could also view the active periods simply as segments of a single, long, fragmented user life. However, we would then expect usage patterns to be continuous between active periods, which is not the case. Instead, we observe that the overall user life is segmented into periods of activity with long inactive periods in between.}.

\begin{figure}[tb]
\vspace{-1mm}
\centering
\includegraphics[width=0.40\textwidth]{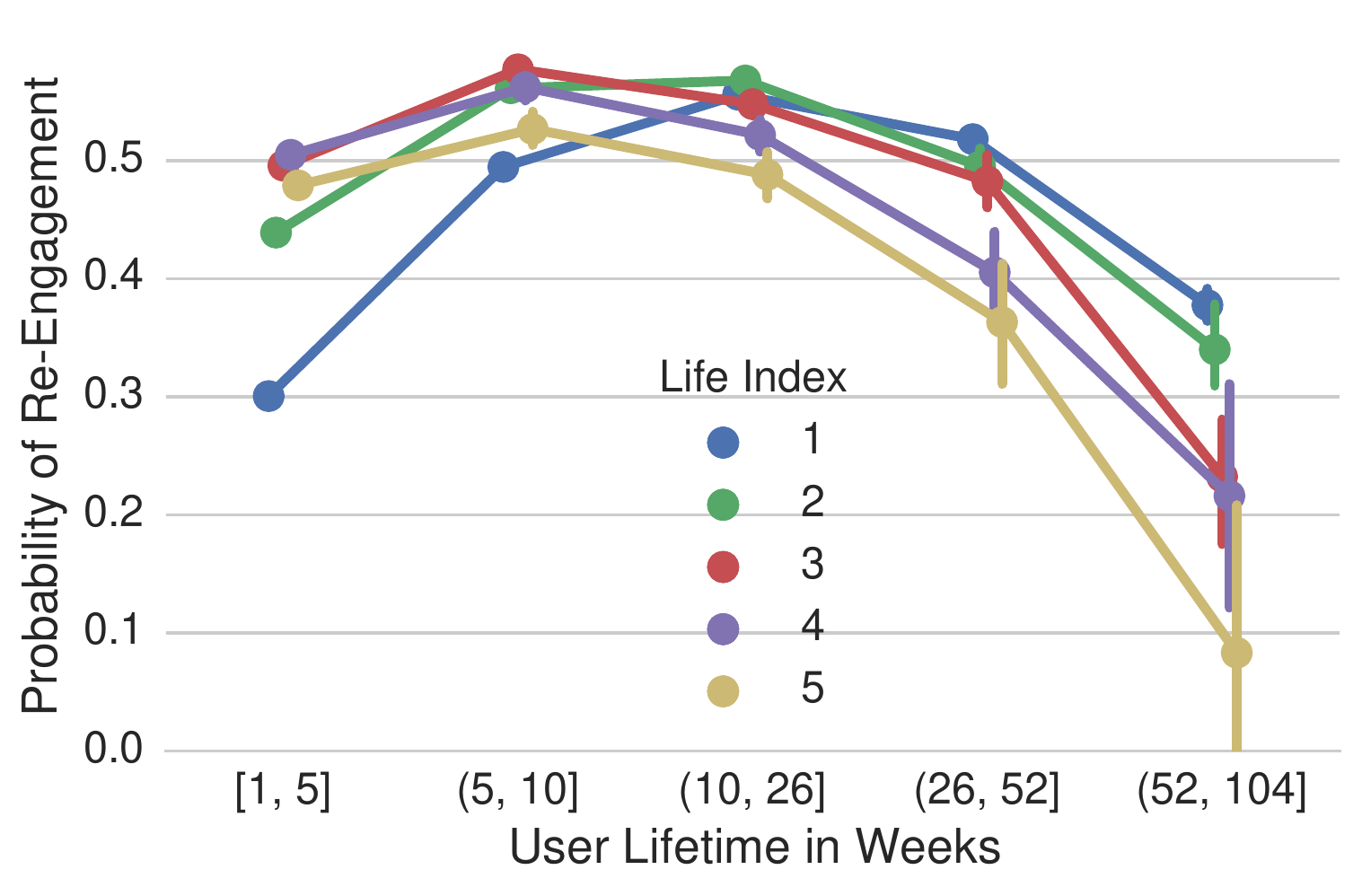}
\vspace{-2mm}
\caption{Probability of users re-engaging in new lives across different user groups of different lifetime.
}
\label{fig:lifetime_vs_prob_coming_back}
\vspace{-2mm}
\end{figure}

\xhdr{Relationship between total user lifetime and the number of lives}
We observe a relationship between the total user lifetime and the number of lives (active periods).
Figure~\ref{fig:lifetime_vs_prob_coming_back} shows the probability of re-engaging for another life as a function of the duration of the previous life.
In the first life (blue curve) we observe a clear U-shape with both very short and very long first lives being associated with a smaller probability of multiple lives.
In later lives, this U-shape relationship attenuates and short lives are associated with the highest rates of return.
Users with a very short \emph{first} lives are unlikely returners because they might not have found the app very useful. However, in later lives, a short life does not mean the same thing---users return more often and have likely found value in the application before.

The existence of multiple life patterns for the vast majority of users shows that while users stop using the app regularly, they do frequently return as well, suggesting that there are times when application is valuable to its users.

\section{Why Multiple Lives?}
\label{sec:why_multiple_lives}

Next, we illuminate why there are multiple lives and what are the mechanisms behind users leaving and returning multiple times?
We demonstrate that engagement patterns across multiple lives vary based on the user's primary intent as well as external influences.

\begin{figure}[t]
\includegraphics[width=.40\textwidth]{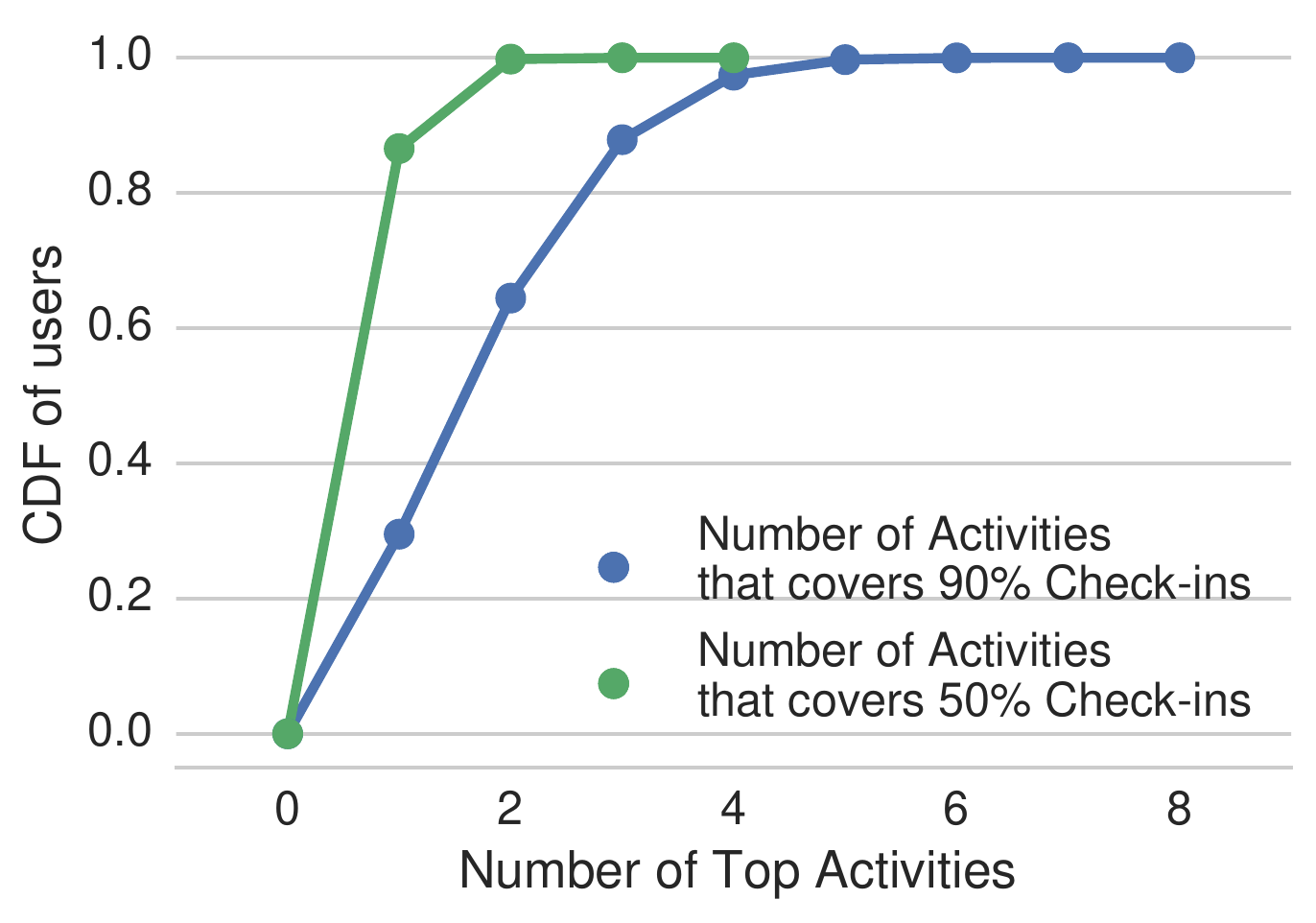}
\caption{Number of activities required to cover a specific fraction of a user's check-ins. 
Users tend to focus on very few activities. For example, one activity already covers 50\% of all check-ins for 87\% of all users (not necessarily the same activity for each user).
}
\label{fig:number_of_activities_took_to_cover_x_checkins}
\end{figure}

\subsection{Multiple User Intents}
\label{subsec:multiple_intents}
Users of an activity tracking application, such as the one studied in this work, use the app with a wide variety of intentions. 
For example, some users might want to lose weight and use the app for regularly tracking their weight changes.
Others might want to be more physically active or sleep better.
We first formalize this notion of user intent and will later show that it helps explain how users are using the app and why their engagement patterns follow multiple lives.

Empirically, we find that most users use the app in a very focused way, only using a small number of different activities. 
Specifically, just one activity is enough to cover 50\% of all check-ins for 87\% of all users (Figure~\ref{fig:number_of_activities_took_to_cover_x_checkins}; green line). 
To cover 90\% of all check-ins, two activities are enough for half the user population (blue line).
Because usage of the app is concentrated on very few activities, we can use the \emph{primary activity} of each user (\ie, the most commonly used one) as a proxy for the user's \emph{primary intent} of using the app.

\begin{figure}[t]
\centering
\vspace{-2mm}
\includegraphics[width=0.40\textwidth]{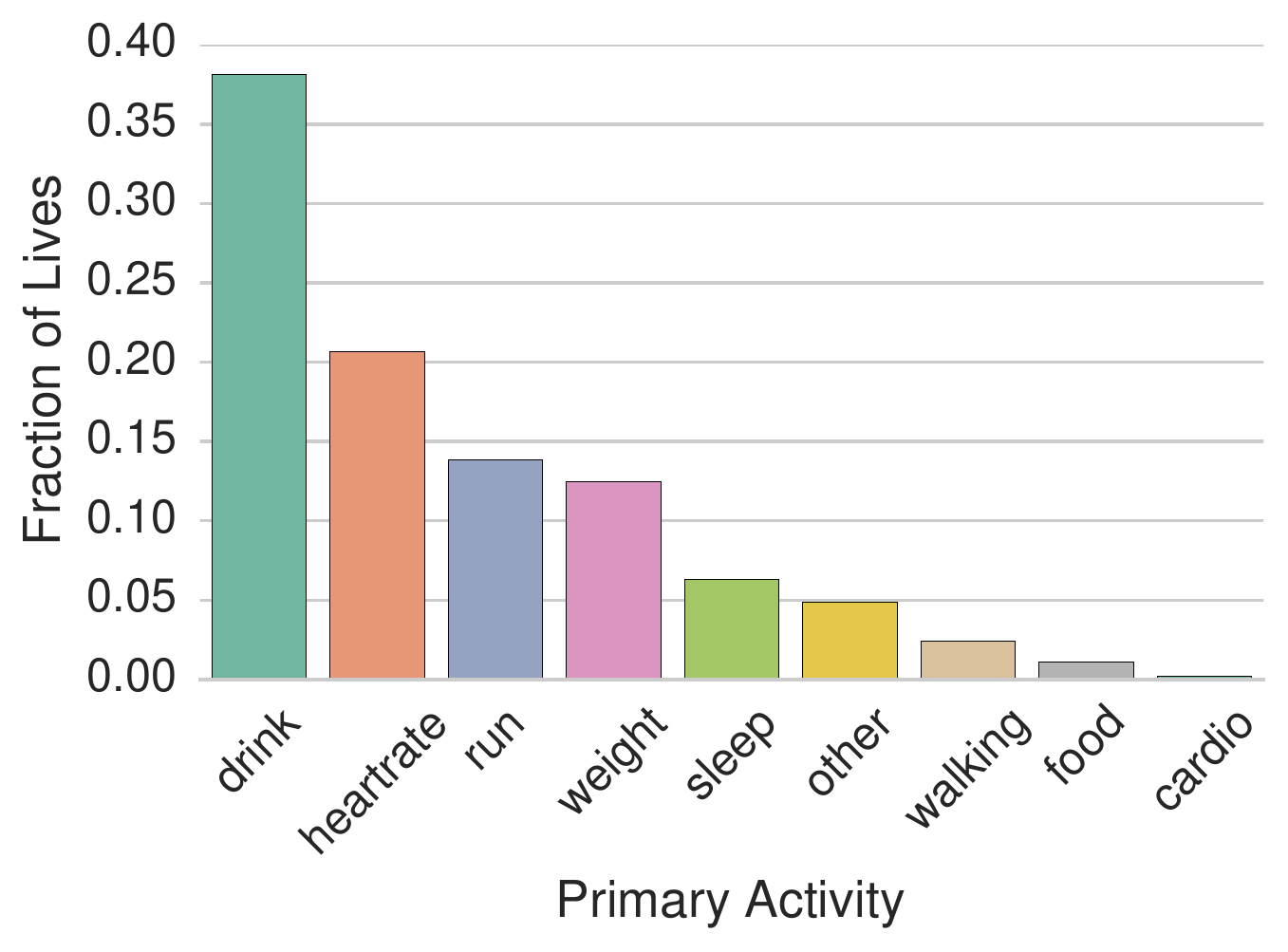}
\vspace{-2mm}
\caption{Fraction of lives focusing on each primary activity. 
Note that users pursue a wide variety of primary activities.
}
\label{fig:primary_activity_portion}
\vspace{-2.5mm}
\end{figure}

The primary activity varies from user to user.
As shown in Figure~\ref{fig:primary_activity_portion}, about one third of users uses the app to track their drinking (\ie, monitor their water, caffeine, or alcohol intake).
One fifth of the users primarily track their heart rate (\eg, immediately after workouts or in a resting condition; both are indicators of cardiovascular health and fitness). 
We observe that other users primarily track their weight (typically with the intent to lose weight), runs, walks, or sleeping patterns. 
A smaller number of users primarily focuses on food logging and cardio activities. 


\begin{figure}[t]
\centering
\includegraphics[width=.40\textwidth]{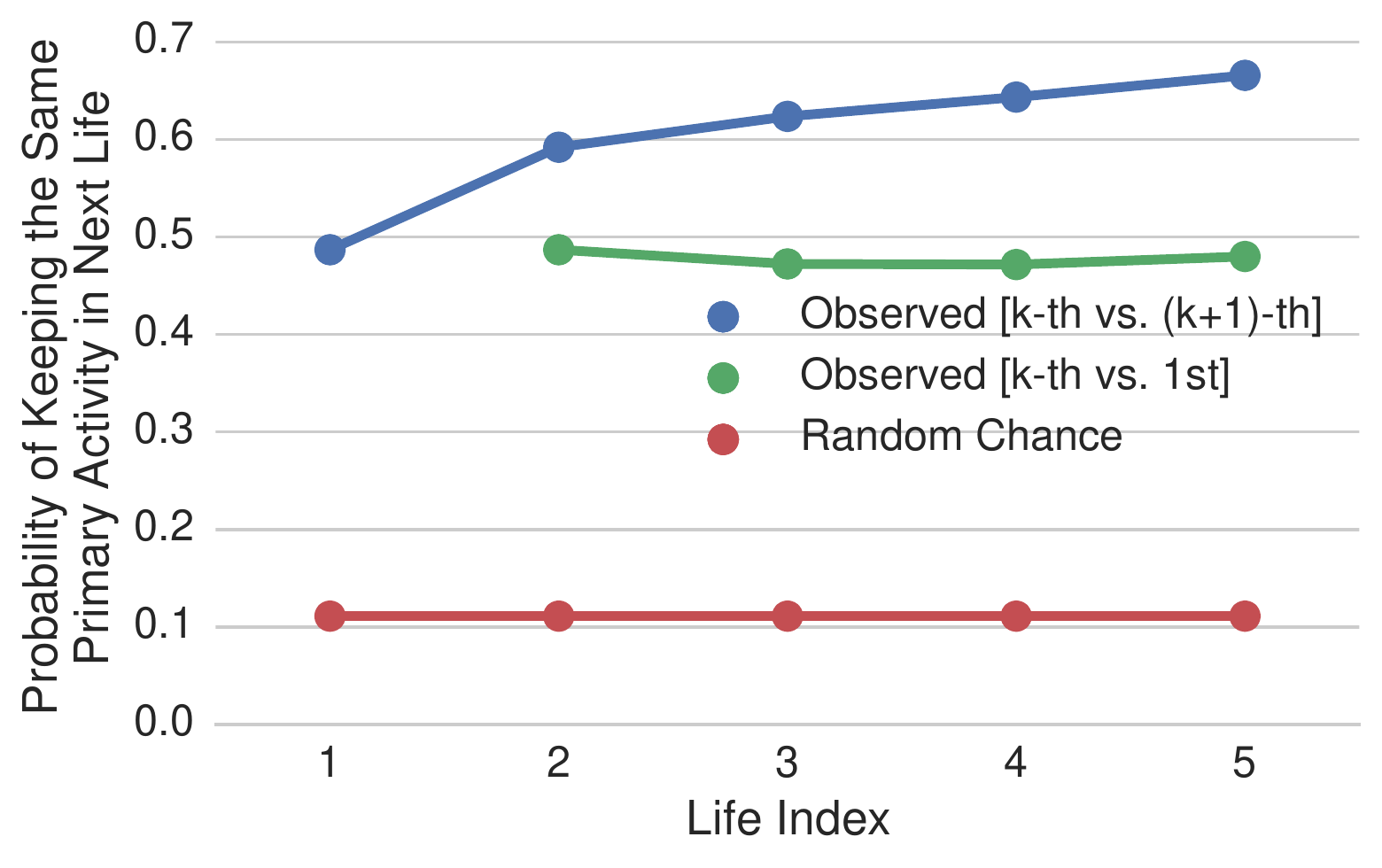}
\caption{
The probability of keeping the same primary activity in the next life ranges from 49\% for the first life to 67\% for the fifth life (blue curve).
Thus, most of the time, users keep the same primary activity across lives. 
The probability of having the same primary activity in the first life ranges from 47\% to 49\% (green curve).
}
\label{fig:same_most_freq_activity}
 \vspace{-2mm}
\end{figure}

\subsection{User Intent \& Multiple Lives}
\label{subsec:intent_multiple_lives}

\xhdrnodot{Do users keep their primary intent in future lives?}
We compute for every user and life which activity a user logged most often (\ie, the primary activity), and how likely a user is to keep their primary activity in consecutive lives.
As shown in Figure~\ref{fig:same_most_freq_activity}, we find that users keep their primary activity from life to life most of the time, ranging from about 50\% after the first life to almost 70\% after the fifth life (blue curve). 
Users' primary activity matches their first primary activity about 50\% of the time (green curve).
These probabilities are substantially higher than consecutive agreement based on picking one of the nine possible activities uniformly at random (red curve).
These findings show that users are typically ``resurrected'' with the same primary intent as before, rather than changing their intentions between consecutive lives.
Note, however, that usage patterns in the next life are not simply a continuation of the previous life but show distinctive signs of a new life as discussed in Section~\ref{subsec:multiple_lives}.

\begin{figure}[t]
\centering
 \vspace{-1.5mm}
\includegraphics[width=0.45\textwidth]{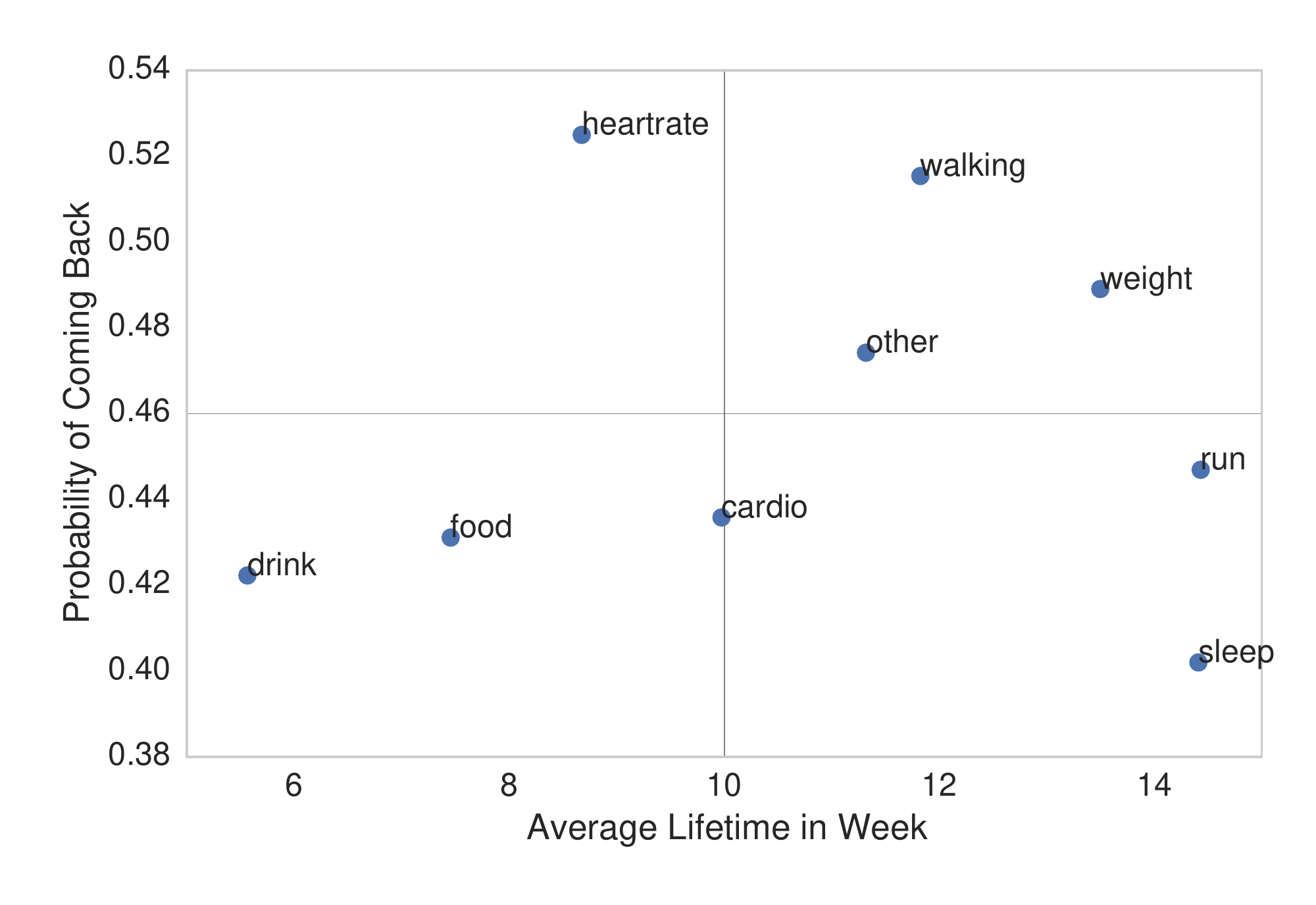}
\vspace{-4mm}
\caption{Average lifetime and return probability vary depending on the user's primary activities. 
}
\label{fig:most_freq_activity_vs_lifetime_vs_prob_of_coming_back}
\vspace{-3mm}
\end{figure}

\xhdr{Multiple life patterns vary with primary intent}
We find that the way users use the app varies based on the user's primary intent. 
Figure~\ref{fig:most_freq_activity_vs_lifetime_vs_prob_of_coming_back} shows the average lifetime in weeks on the x-axis and the probability of returning for another life on the y-axis across primary intents of the user (here, we only consider user lives with at least 10 check-ins to reduce noise).
Empirically, we find that users primarily logging their heart rate are most likely to re-engage for another life (52\%). 
Users who primarily log runs and sleep are less likely to return, but their average lifetime is almost twice as long compared to primary heart rate users (14.4, 14.4 vs 8.7 weeks). 
This suggests that heart rate measurements are used every now and then, but that most users do not monitor them over long periods of time.
In contrast, run- and sleep-loggers appear to find the associated features (GPS trail, timing and pace statistics; sleep timing, duration and quality) useful over longer periods of time. 
Activities such as drink- and food-logging are associated with both few and short lives.
On the other hand, walking- and weight-logging activities are linked to more and longer lives, suggesting that users regularly find value in this functionality and for long periods of time. 

We discuss design implications for increasing user engagement based on this heterogeneity in multiple life patterns based on primary intent in Section~\ref{sec:implications_user_engagement}.


\begin{figure}[t]
\begin{subfigure}{.23\textwidth}
  \includegraphics[width=\textwidth]{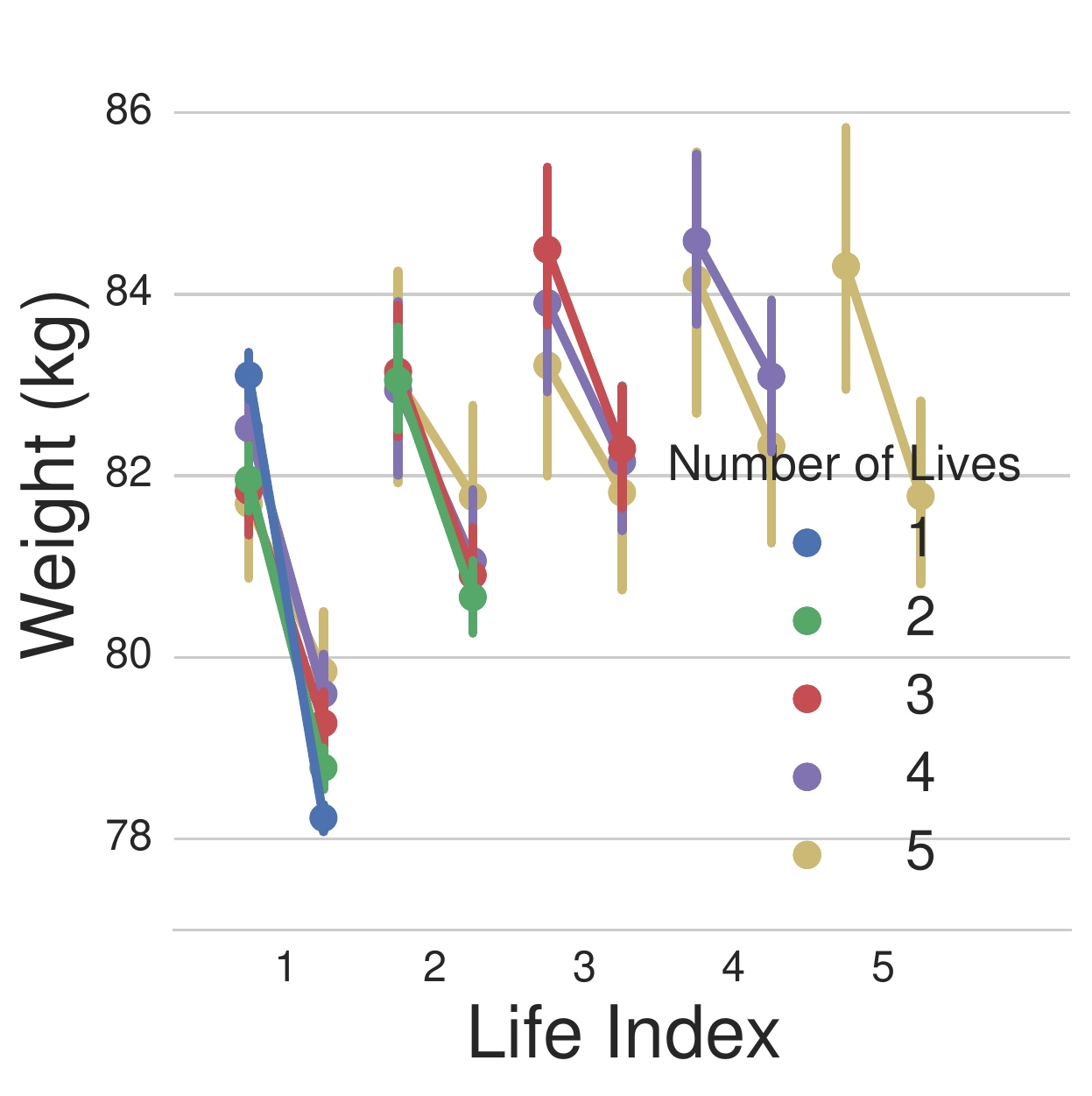}
  \vspace{-6mm}
  \caption{First and last logged weight of each life}
     \label{fig:zig_zag_weight}
\end{subfigure}
\hspace{0.8mm}
\begin{subfigure}{.215\textwidth}
  \vspace{2mm}
  \centering
  \includegraphics[width=\textwidth]{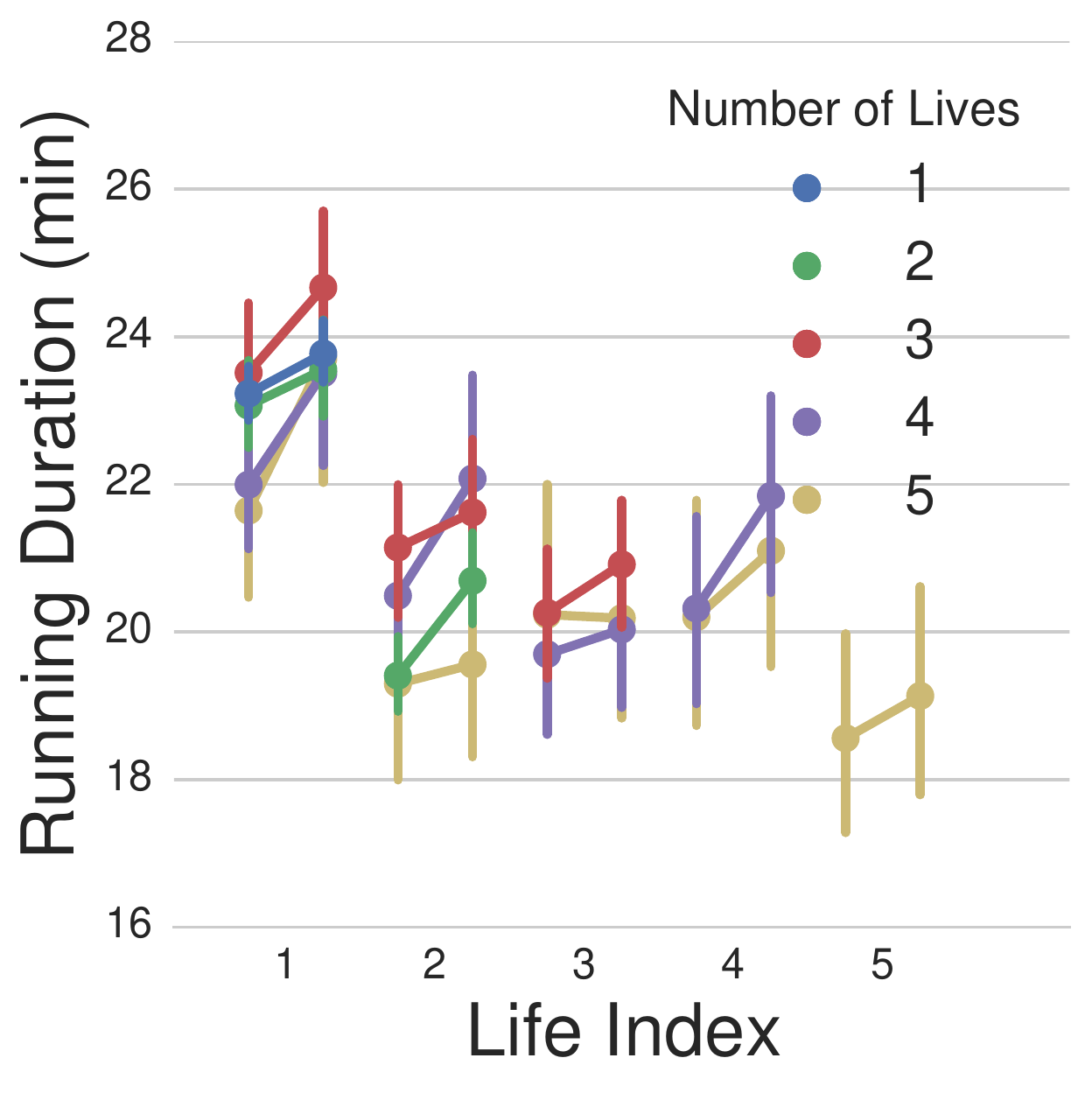}
  \vspace{-6mm}
  \caption{First and last logged running duration of each life.}
     \label{fig:zig_zag_running}
\end{subfigure}
\vfill
\begin{subfigure}{.23\textwidth}
  \centering
  \includegraphics[width=\textwidth]{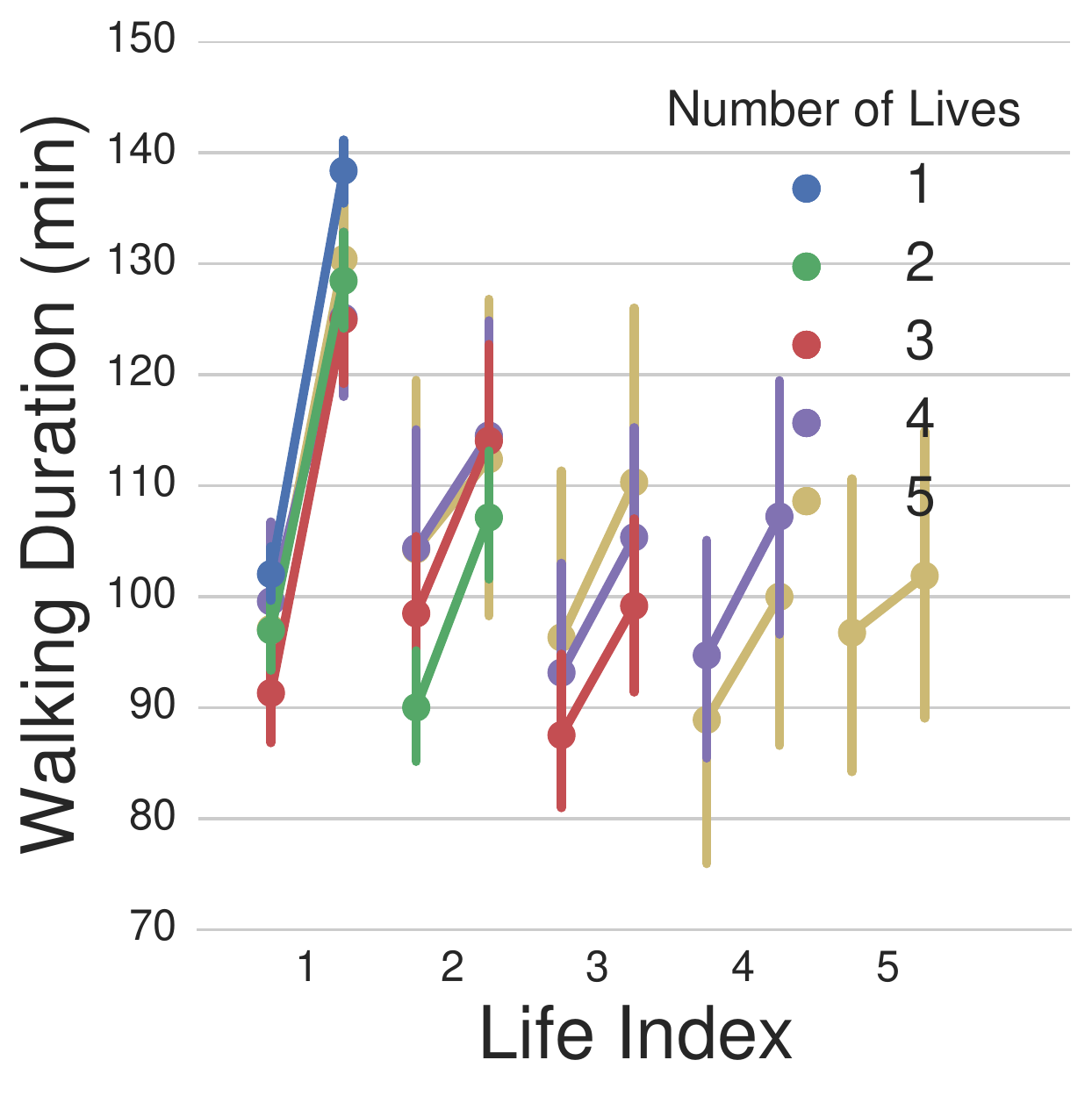}
  \vspace{-6mm}
  \caption{First and last logged walking duration of each life.}
     \label{fig:zig_zag_walking}
\end{subfigure}
\hspace{0.8mm}
\begin{subfigure}{.23\textwidth}
  \centering
  \includegraphics[width=\textwidth]{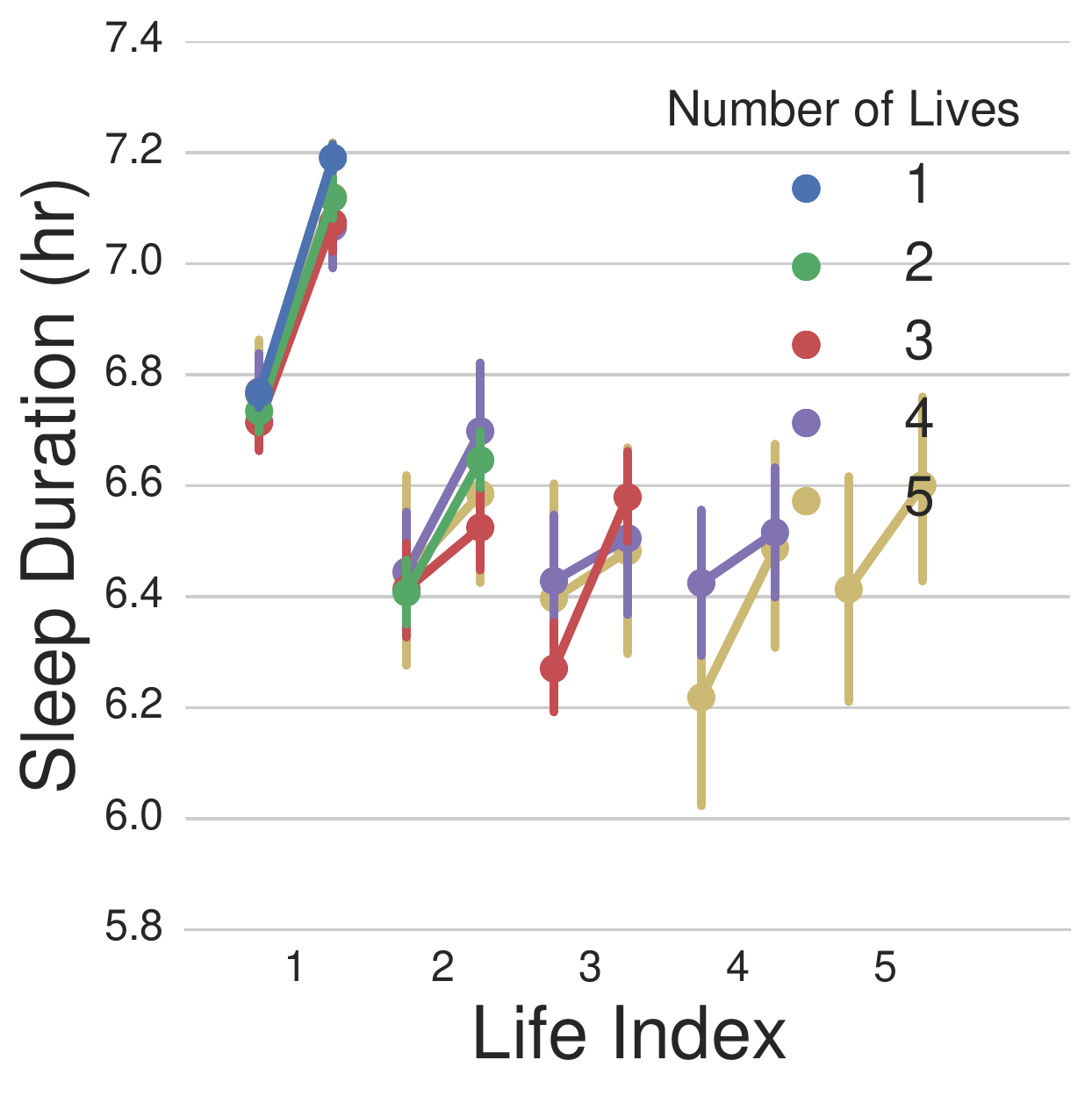}
  \vspace{-6mm}
  \caption{First and last logged sleep duration of each life.}
     \label{fig:zig_zag_sleeping}
\end{subfigure}
\vspace{-1mm}
\caption{Average first and last logged value of weight (a), running (b), walking (c), and sleep duration (d).
We find that users have lower weight, and longer running, walking, and sleep durations when they are about to stop using the app.
However, when they re-engage, they log a higher weight and shorter running, walking, and sleep durations.
}
\label{fig:performance_zig_zag_plots}
\vspace{-3mm}
\end{figure}

\subsection{Intent-oriented Performance Driving Multiple Lives}
\label{subsec:intent_performance}

Associated with the user's primary intent are specific performance goals.
For example, a user logging weight would often be interested in losing weight, a user logging running or walking would typically be interested in doing so longer or faster, and users logging their sleep would often be interested in sleeping longer or more consistently.
We hypothesize that users may stop using the application once they significantly improved towards a goal associated with their primary intent. 
Furthermore, these users may return to the app once they have a new goal to attain.
In order to test this hypothesis, we choose four quantities for which we have associated performance outcomes that are reported through the application (Figure~\ref{fig:performance_zig_zag_plots}): 
weight activity using weight loss in kilograms, running activity using run duration in minutes, walking activity using walk duration in minutes, and sleep activity using sleep duration in hours (higher is better, except for weight as most users are overweight or obese; Table~\ref{tab:dataset}).
We measure these activity-associated outcomes for both the first and last logged activity in each life. 
Furthermore, we consider user populations with a different number of lives separately in order to compare unchanging populations over time. 

\xhdr{Results}
We show average activity-associated outcomes for weight (Figure~\ref{fig:zig_zag_weight}), running (Figure~\ref{fig:zig_zag_running}), walking (Figure~\ref{fig:zig_zag_walking}), and sleep activities (Figure~\ref{fig:zig_zag_sleeping}) with user groups of different number of total lives in different colors. 
Across all four plots we observe a ``zig-zag'' pattern, where users start their new life with a lower level of performance than they end it with; that is, weight decreases, runs duration increases, walk duration increases, and sleep duration increases within each life.
Furthermore, we find that in all cases when users are returning to the app, they have a lower performance level again (\eg, regained weight, or shorter runs/walks/sleeping). 
We find very similar result when focusing on primary users of each activity only.

This suggests that making significant progress towards these performance goals is associated with leaving the app, in which case the app might have fulfilled its purpose for that particular user.
However, users may set themselves new goals or experience setbacks (\eg, increased weight or decreased fitness levels) giving them a reason to return to the activity tracking app, which may have helped them previously to realize their goals. 
Then, the app becomes useful again to improve performance in some domain.
We note that the ``zig-zag'' patterns in Figure~\ref{fig:performance_zig_zag_plots}, where users leave the app with improved performance and return with decreased performance, 
provide additional evidence that user engagement in this activity tracking application is indeed segmented into multiple lives.

\begin{table}[t]
\centering
\resizebox{1.0\columnwidth}{!}{%
\begin{tabular}{lcccc}
\toprule
Outcome\textbackslash Prim. Activity & Weight & Running & Walking & Sleep \\ 
\midrule
Weight Change (kg) & -7.5*** & -3.9*** & -5.2*** & -1.4***  \\
Running Duration (min) & 1.3 & 0.3*** & 2.00* & 0.0 \\
Walking Duration (min) & 23.7*** & 21.8*** & 42.0*** & 22.8*** \\
Sleep Duration (hr) & 0.4*** & 0.7*** & 0.5** & 0.2*** \\
\bottomrule
\end{tabular}
}
\vspace{1mm}
\caption{Users that focus on a particular activity, tend to improve outcomes associated with that activity (\eg, primarily-walkers increase their walk durations but also tend to lose weight). 
Every column is a different user population with corresponding primary activity. 
Rows correspond to different outcome measures. 
Wilcoxon signed-rank test is used to test significance (* $p<0.05$, ** $p<0.01$, ***$p<0.001$). 
} 
\label{table:primary_activity_effect}
\vspace{-5mm}
\end{table}

\xhdr{Users typically improve performance of primary activity}\\ 
Here we provide additional evidence that users focusing on a certain activity tend to improve performance outcomes associated with that activity (Table~\ref{table:primary_activity_effect}). 
Specifically, users tend to improve the most on their primary activities. 
For example, primary-walkers improve the most (42 minutes) on walking durations.
For users primarily logging weight, we find that the last weight logged by these users is on average 7.5 kilograms lower than their first logged weight. 
These users lost the most weight compared to other groups focusing on different primary activities.
We note that activities are also intimately related.
For example, primarily logging weight is not the only way app users can lose weight.
In fact, we find that primarily-walkers and primarily-runners also log significant weight loss (-5.2kg and -3.9kg, respectively) and increased sleep duration (0.5 and 0.7 hours, respectively).
This highlights the complexity of interrelated activities and health behaviors, where some activities may replace another, while other activities may even further support each other and the goals associated with them (\eg, increased physical activity from running and walking may support weight loss). 
Overall, we find that users tend to improve performance outcomes associated with their primary intent, and that users may leave the app afterwards.

\begin{figure}[t]
\centering
\vspace{-2mm}
\includegraphics[width=0.37\textwidth]{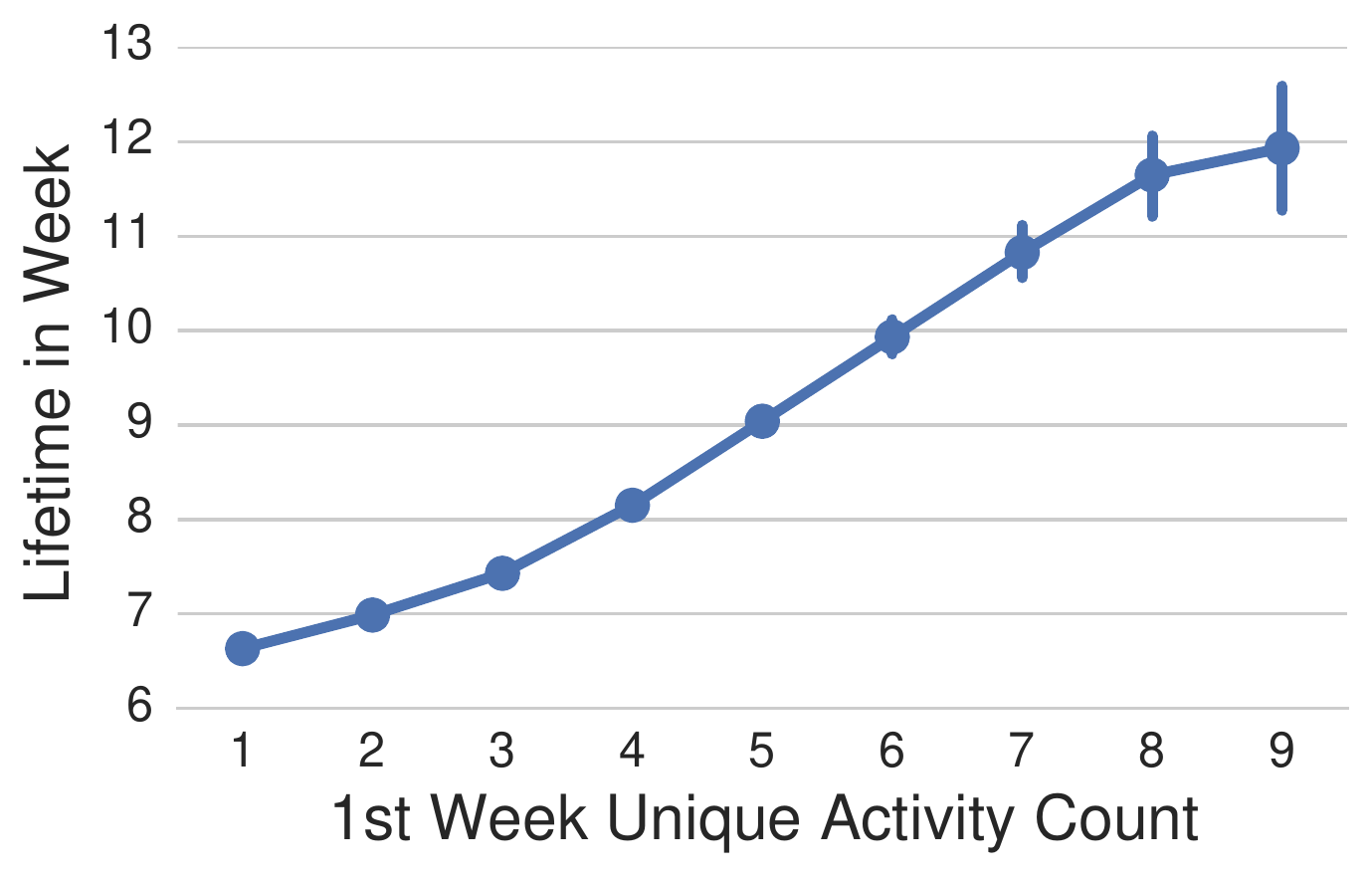}
\vspace{-2mm}
\caption{Users with higher number of distinct activities during the first week experience longer lifetimes.}
\label{fig:unique_count_vs_lifetime}
\end{figure}

\xhdr{Users with more diverse activities live longer}
We just observed that, for example, users primarily focusing on running or walking also logged large weight loss. 
Intuitively, this makes sense.
In order to lose weight, one needs to do more than just log weight. 
Instead, one needs to increase energy expenditure through exercise or reduce energy intake from food and drinks. 
Logging multiple activities could be a marker of increased interest in one's health or a marker of intended behavioral changes that would support specific health goals. 
Here, we test the hypothesis whether users with more \emph{diverse} sets of activities (\eg, running and weight instead of just weight logging activities) would be different from less diverse users in terms of their lifetime duration.
%
Empirically, we find that users with more diverse initial usage patterns live longer (Figure~\ref{fig:unique_count_vs_lifetime}).

\begin{figure}[t]
\centering
\vspace{-3mm}
\includegraphics[width=0.37\textwidth]{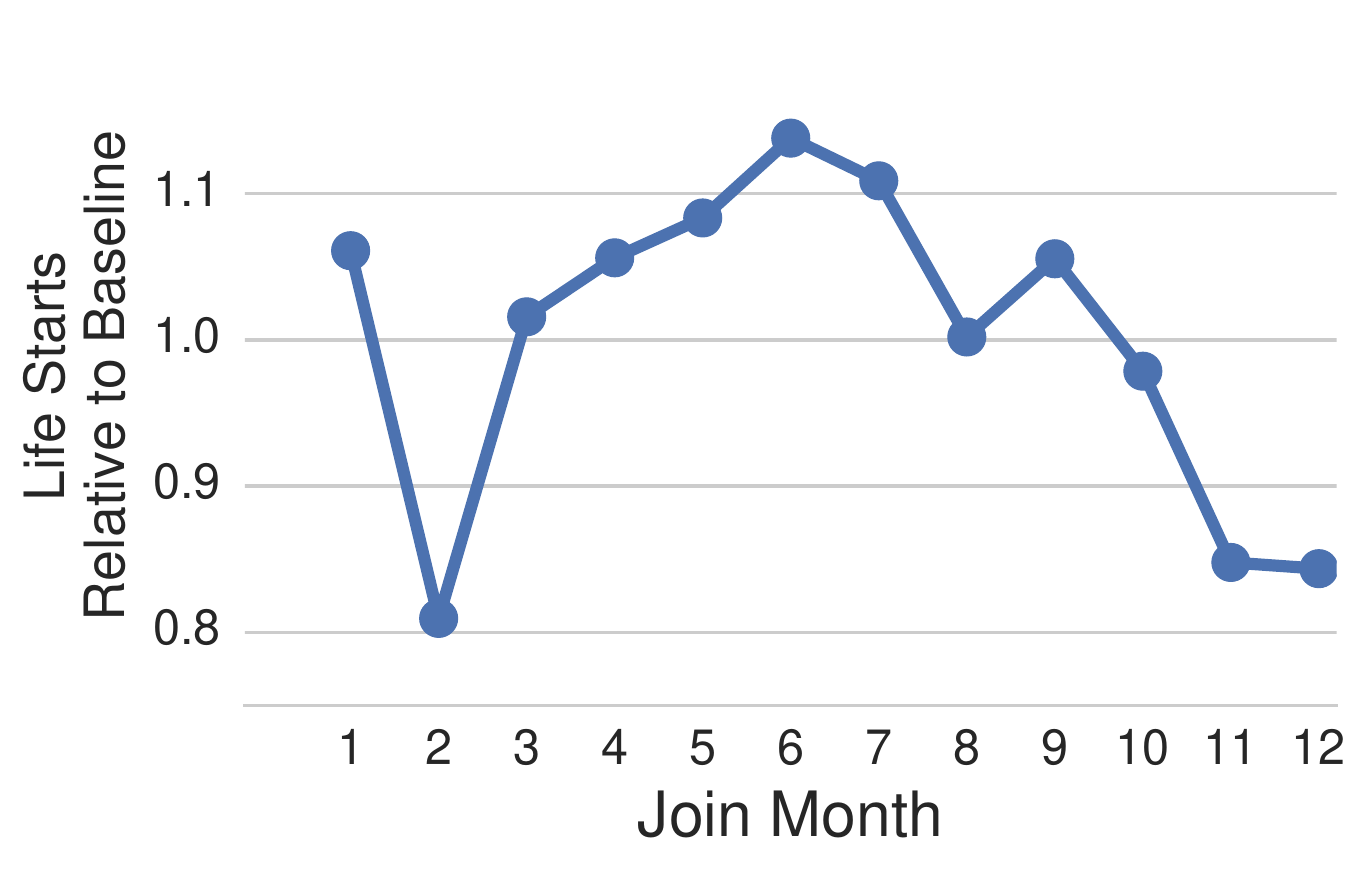}
\vspace{-2mm}
\caption{Users start new lives predominantly in January (6\% over average rate) and summer months (14\% over average rate), possibly related to New Year's resolutions and summer related activity and weight loss goals.
}
\label{fig:join_month}
\vspace{-3mm}
\end{figure}

\subsection{External Influences}
\label{subsec:external_influences}
The activities tracked by activity tracking applications take place in the real, physical world. 
As such, they are influenced by the practices and demands of the real world. 
For example, there may be seasonal influences driving user intents and goals.
To investigate external, seasonal influences, we consider each new user life and count the number of lives that are started in each month of the year (\ie, users either start using the app for the first time or return to the app after a prolonged inactivity).
We normalize with respect to the average monthly rate. 
We only consider observations between January 2014 and December 2015 to ensure that our observation period includes exactly two instances of each month.

\xhdr{Results}
We observe that the distribution of starting time of new lives is clearly non-uniform (Figure~\ref{fig:join_month}).
We find the highest number of new lives starting in January (6\% over average rate) and June (14\% over average rate). 
These findings can be explained by external influences driving user intentions.
For example, most users come from the United States and other western, developed countries in the Northern Hemisphere.
In these countries, New Year's resolutions are commonly expressed on January 1; a tradition in which a person resolves to change an undesired trait or behavior, to accomplish a personal goal, or otherwise improve their life. 
Many New Year's resolutions revolve around being more physically active or losing weight and users with these resolutions may seek out activity tracking applications to support them.
Furthermore, people often seek to lose weight or be more active during the summer months around June. 
During this time weather is often more favorable to physical activity and users may attempt to lose weight for the ``swimsuit season''. 
The patterns of when new lives are started suggest that users of activity tracking applications are externally influenced by real-world practices and demands.
Particularly in this example, user (re-)joining time may be influenced by seasonal effects.

We note that multiple lives patterns also vary with demographic factors such as age, gender, and weight status. 
For example, we find that young users have much shorter lifetimes and are less likely to re-engage after long inactivity than older users. 
We will exploit these correlations in Section~\ref{sec:prediction_task} in order to predict whether a user will re-engage in a new life and to predict their lifetime.

\section{Predicting Multiple Lives}
\label{sec:prediction_task}


We formulate a novel prediction task of predicting how many lives a user will have.
This section leverages previously described insights, including a marker of improved primary intent performance, in order to predict whether a user comes back for another life.
We also predict how long a given life will last.
We demonstrate that while there is large variability in user re-engagement patterns, the factors studied in this work allow us to successfully predict whether a user re-engages for another life (71\% ROC AUC) and how long this life will last (82\% ROC AUC).
We note that the prediction tasks are designed to validate our empirical findings. 
Features derived from our findings are potentially generalizable to capture multiple lives in other similar applications.

\subsection{Task Description}
\xhdr{Future Life Prediction} 
For any (completed) user life, we predict whether this user will have additional future lives; that is, whether they will re-engage with the app again after prolonged inactivity.
This is a new prediction task based on the multiple life paradigm introduced in this work.

\xhdr{Lifetime Prediction}
For any user life, after observing the first four weeks ($w=28$ days), we predict if a user is going to be a \emph{short-term user} (leaving within the next $m=30$ days) or a \emph{long-term user} (living longer than $n=183$ days).
This follows the setting proposed by Danescu-Niculescu-Mizil et al.~\cite{danescu2013no}.
In this setting, we drop users with lifetime in $[w+m,n)=[58,183)$ days to increase contrast between the classes. 
We also evaluated our prediction models for different parameter choices and found that our results, including the relative predictive power of individual features, were robust across a wide range of parameters.

\subsection{Experimental Setup}
Note that both tasks are formulated as binary prediction tasks. 
To define the user lives, we use $\delta=30$ as before (again, we find very similar results for other choices of $\delta$).
In order to avoid pre-mature classification of whether users will re-engage or not, we only consider user lives that are complete well before the end of the observation period.
Specifically, we ensure that all considered users have no check-ins within the last 180 days of the observation period (this is larger than the average inter-life gap which is 114 days after the first life and shorter for following lives; see Figure~\ref{fig:inter_life_gap}). 
Note that by the very nature of the re-engagement patterns described in this work, one can never be sure whether a particular users may still re-engage after 180 days. 
However, we tested various other thresholds and found similar results. 
In total, this leaves us with 1,267,897 user lives of 851,582 distinct users for prediction.
We use the area under the ROC curve (ROC AUC) as our evaluation metric and 
use 10-fold cross validation for estimation. 
We report performance for Gradient Boosted Tree models and optimize number of trees, tree depth, and learning rate through cross-validation on the training data. We also experimented with Logistic Regression and
linear SVM models, which consistently gave lower performance due to prominent nonlinear relationships (\eg, users with medium-length lifetime are most likely to return to the app; Figure~\ref{fig:lifetime_vs_prob_coming_back}).


\xhdr{Models}
We define a series of models with different feature sets in order to learn what features are most predictive of future lives and lifetime. 
We use the same features for both prediction tasks. 
However, in lifetime prediction, we compute these features from only the first four weeks of each life. 
If features are missing we impute zero and include a binary variable indicating missingness.

\begin{enumerate}
  \item \textbf{Lifetime:} Lifetime of current life in days (Section~\ref{subsec:multiple_lives}).
  (Of course, we exclude this feature when predicting lifetime.) 
  \item \textbf{Usage Pattern:} Weekly numbers of check-ins, number of distinct activities, entropy of activity distribution, and fraction of check-ins from most frequent activity (Section~\ref{subsec:multiple_lives}). We also include the week-to-week changes in these metrics as features.
  \item \textbf{Primary Activity:} Categorical variable indicating the user's most frequent activity (Section~\ref{subsec:intent_multiple_lives}). 
  \item \textbf{Performance Change:} 
  As a marker of improved primary intent performance, we include the change in weight, and running, walking, and sleep durations as well as the number of check-ins for each activity (Section~\ref{subsec:intent_performance}).
  \item \textbf{Demographics:} Three categorical variables for age, gender, and body-mass index (Section~\ref{subsec:external_influences}).
  \item \textbf{Join Time:} Number of days between the user's first activity and the launch of Argus app in July 2013 (Section~\ref{subsec:external_influences}).
  \item \textbf{All:} A combination of all features. 
\end{enumerate}



\begin{figure}[t]
\centering
\vspace{-1mm}
\includegraphics[width=0.49\textwidth]{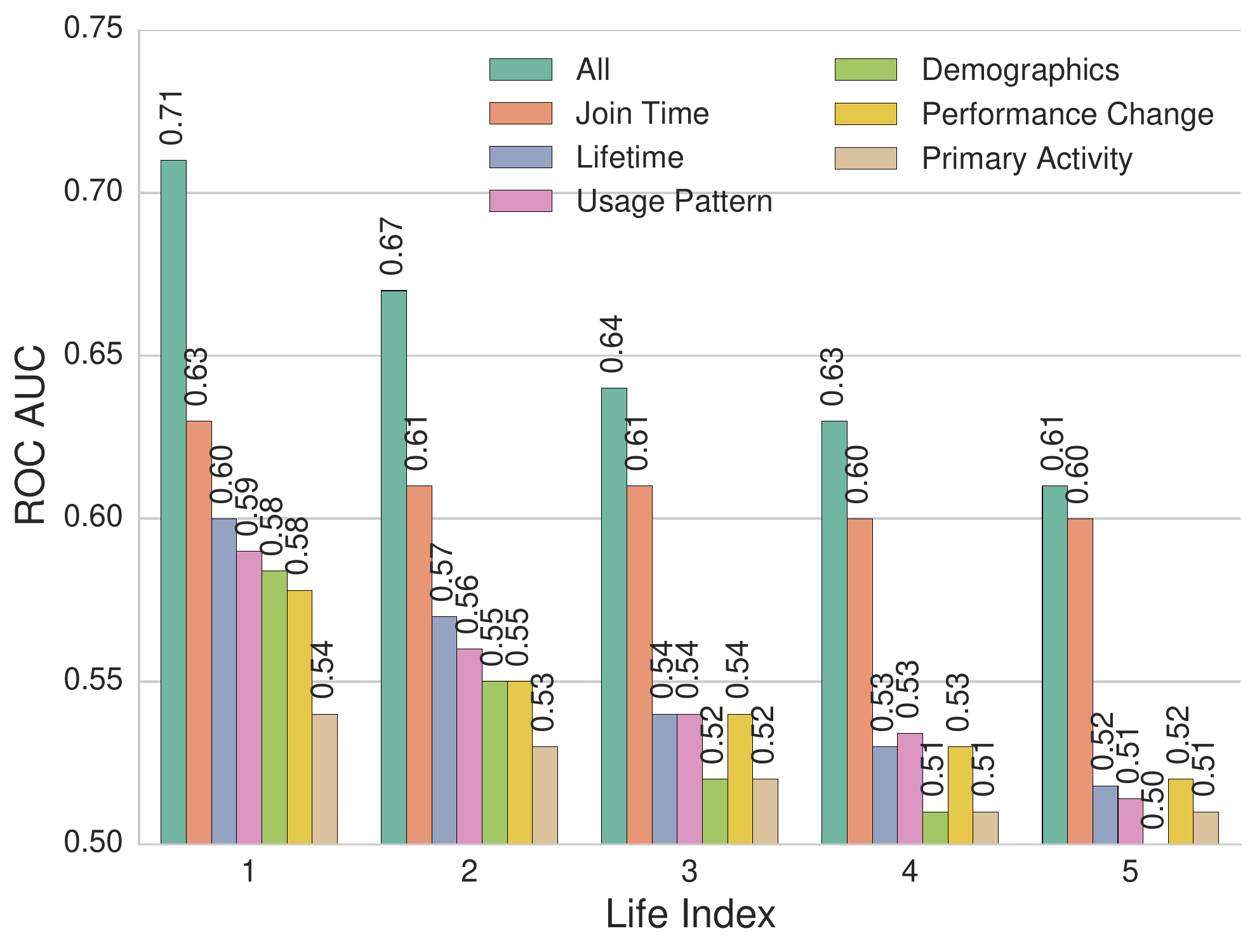}
\vspace{-6mm}
\caption{
Future life prediction performance. We predict at the end of each life whether a user will re-engage eventually. 
}
\label{fig:come_back_prediction_feature_comparison}
\vspace{-2.5mm}
\end{figure}


\begin{figure}[t]
\centering
\includegraphics[width=.49\textwidth]{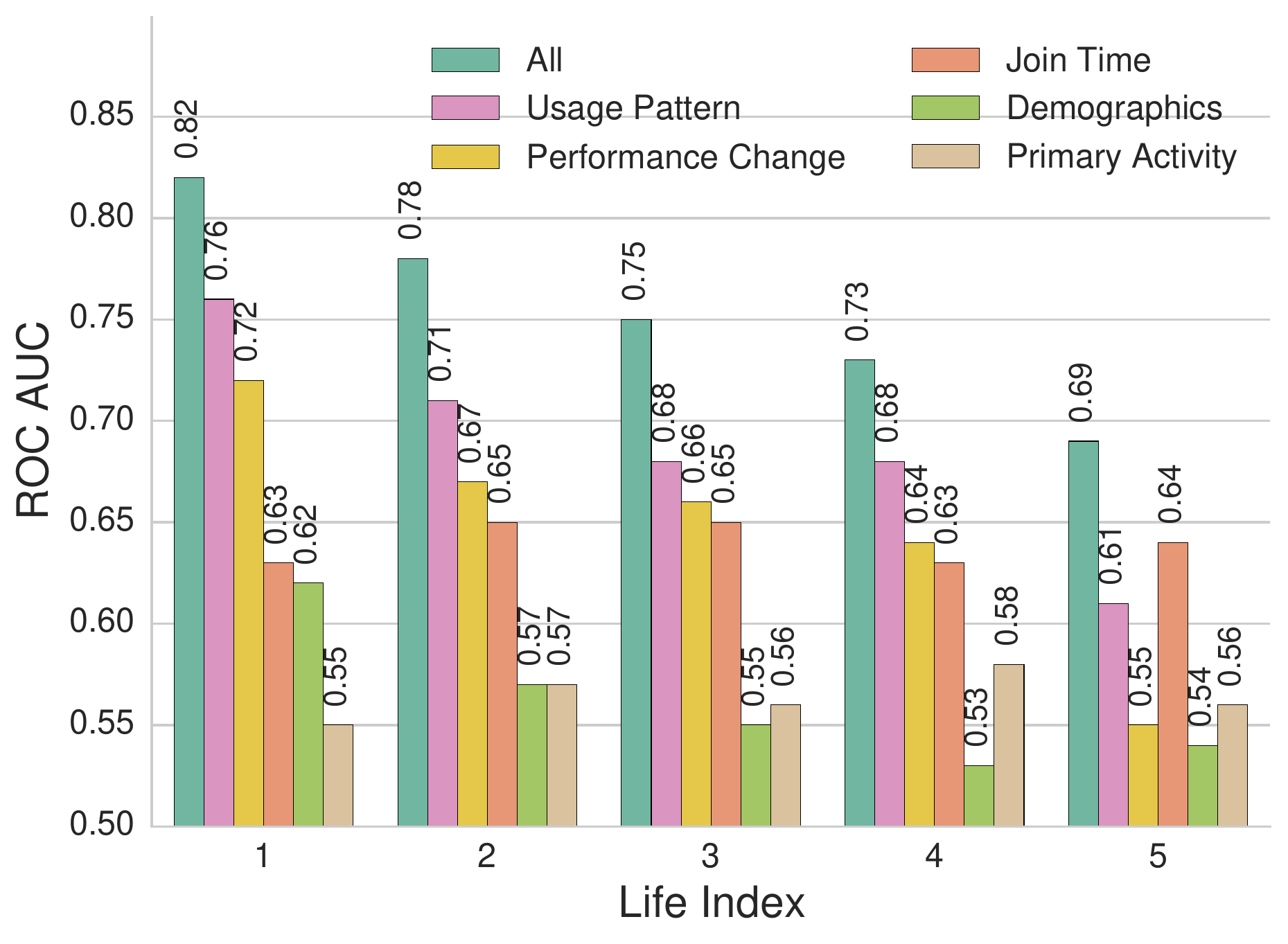}
\vspace{-6mm}
\caption{Lifetime prediction performance. 
We predict whether a user will leave shortly or stay long-term. 
} 
\label{fig:lifetime_prediction_feature_comparison}
\vspace{-2.45mm}
\end{figure}

\subsection{Results}
\xhdr{Future Life Prediction} 
The prediction accuracies for predicting whether a user will re-engage with the app in the future are shown in Figure~\ref{fig:come_back_prediction_feature_comparison}, separately for the first to fifth life of each user.
For predicting whether a user will re-engage after their first life, our full model using all features achieves 71\% ROC AUC.
Note that a random baseline would achieve 0.50 ROC AUC.
We observe that all feature groups carry significant predictive performance. 
In addition, we find that the performance change marker (Section~\ref{subsec:intent_performance}) predicts user re-engagement with up to 58\% ROC AUC. 
Further, prediction performance drops with each life. 

\xhdr{Lifetime Prediction}
The prediction accuracies for lifetime prediction are shown in Figure~\ref{fig:lifetime_prediction_feature_comparison}, again separately for the first to fifth life of each user.
Our full model achieves 82\% ROC AUC for the first life. 
We again observe that all feature groups carry significant predictive performance. 
We find that the performance change marker is more predictive of lifetime than re-engagement with up to 72\% ROC AUC (compared to 58\% ROC AUC before).
Again, we observe that the task becomes harder for later lives with full model accuracy ranging from 82\% to 69\% ROC AUC for the fifth life.

 \xhdr{Discussion}
In the lifetime prediction task, usage patterns and performance changes are the most predictive features to distinguish long-term users from short-term users, as highly engaged users with frequent check-ins also tend to stay longer using the app (Section~\ref{subsec:intent_performance}).
In the future life prediction task, the join time is highly predictive of re-engagement since it differentiates early adopters from late-joining users as well as identifying users joining in specific months including those with New Year's resolutions and summer weight loss goals (Section~\ref{subsec:external_influences}).
Prediction performance drops in later lives is potentially due to both fewer data points in later lives (Figure~\ref{fig:ccdf_number_of_lives_dist}) and more noise due to shorter lifetime (Figure~\ref{fig:lifetime_dist}). 
Users in earlier lives display greater changes in usage pattern (Figure~\ref{fig:zig_zag_summary_user_behavior}) and performance (Figure~\ref{fig:performance_zig_zag_plots}), which may also contribute to better prediction accuracy in earlier lives as they are more distinguishable from noise in data.
Our results demonstrate that the insights described in this work allow to successfully predict whether a user re-engages for another life and how long this life will last. 
These models could be used to identify well-suited user populations that could be targeted with additional notifications, e-mails and incentives, with important implications for increasing user engagement and retention.





\section{Related Work}
\label{sec:related}

\xhdr{User engagement}
User engagement has been defined as the ``quality of the user experience that emphasizes the positive aspects of interacting with an online application and, in particular, the desire to use that application longer and repeatedly~\cite{lalmas2014measuring}.''
An extensive literature has studied how to measure user engagement through subjective (\eg, self-report) and objective measures (\eg, eye tracking, mouse movements)~\cite{attfield2011towards}.
In the context of mobile applications, engagement has been measured via clicks on notifications, time past before seeing the notification, and usage time~\cite{mehrotra2017understanding}.
In this work, we used simple, objective statistics to operationalize user engagement based on the frequency of interactions with the application.
Previous work on user engagement has discussed the concept of \emph{re-engagement} where users interrupt usage for a few minutes or hours~\cite{obrien2008user,obrien2011exploring}. Many studies were conducted over a brief period of time which made studying re-engagement patterns impossible.
In contrast to brief usage interruptions, in this work we study re-engagement patterns after extensive inactivity (\eg, 30-90 days).

\xhdr{Modeling \& predicting user engagement}
User engagement has previously been modeled based on repeat consumption patterns~\cite{anderson2014dynamics,benson2016modeling}, binge watching~\cite{trouleau2016just}, and patterns of switching between boredom and sensitization~\cite{kapoor2015just}.
A user's intent in using a particular application can be predicted from the user's behavior~\cite{lo2016understanding,cheng2017predicting} 
since it is strongly influenced by their intent~\cite{ajzen1985intentions}.
Typical modeling of user engagement has considered users to have a single ``lifetime'' during which the user typically becomes less and less engaged on the platform~\cite{yang2010activity,danescu2013no}. 
Much research has been devoted to predicting time of next user activity~\cite{althoff2015donor,kapoor2014hazard,jing2017neural}, 
total lifetime~\cite{danescu2013no,wang2016accurate}, often using survival modeling techniques~\cite{dave2017fast,kapoor2014hazard,jing2017neural}.
In the context of wearable and mobile device data, user engagement research has attempted to proactively engage users~\cite{pielot2017beyond}, infer the mood of users~\cite{likamwa2013moodscope}, and suggest healthy behavior changes~\cite{rabbi2015mybehavior,hansel2015challenges,hansel2015wearable}.
All these studies assumed that user engagement follows a single life paradigm.

\xhdr{Increasing user engagement}
Previous studies have shown that user engagement can be increased through notifications~\cite{alkhaldi2016effectiveness}, incentives including badges~\cite{anderson2014engaging,anderson2013steering}, gamification~\cite{hamari2014does,althoff2016influence,hansel2015wearable}, and social network features~\cite{althoff2017onlineactions,hansel2015wearable}.
Furthermore, understanding the user's intent~\cite{lo2016understanding,cheng2017predicting} can help designers surface different interaction modes~\cite{sherwin2016intentdesign}, provide better contextual help~\cite{stumpf2005predicting}, and personalize search results and recommendations~\cite{teevan2008personalize,teevan2005personalizing}.

\xhdr{This work}
This work studies user engagement and re-engagement patterns after long periods of inactivity in the context of activity tracking applications. 
We extend previous work (\eg,~\cite{shameli2017gamification,althoff2017onlineactions,althoff2017large,ahtinen2008tracking,rooksby2014personal,lister2014just,west2012there}) by discovering that users regularly re-engage after long periods of inactivity structuring user engagement into \emph{multiple lives} with distinct characteristics.
We demonstrate that these multiple lives are driven by user intent and external influences. 
Further, we propose a novel prediction task of predicting the number of lives of a user.

\section{Discussion \& Conclusion}
\label{sec:implications_user_engagement}

Finally, we summarize our findings in the form of a conceptual model of user engagement across multiple lives.
We also discuss design implications for increasing user engagement in this setting.

\subsection{Conceptual Model of Multiple Life User Engagement in Activity Tracking Apps}
The presented empirical evidence can be explained by the following model of user engagement across multiple lives:
\begin{enumerate}
  \item Users join the application with a specific, primary intent and associated goals, starting a new life (Section~\ref{subsec:multiple_intents}).
  \item The user uses the app in accordance with their primary intent (Section~\ref{subsec:multiple_intents}). Users typically improve in performance metrics associated with their primary intent during this time (Section~\ref{subsec:intent_performance}).
  \item Once the user have made significant improvements towards their goal they may leave the app, corresponding to the end of their lifetime (Section~\ref{subsec:intent_performance}). 
  \item When the user forms a new goal (often based on external, seasonal influences; Section~\ref{subsec:external_influences}), or experiences a setback (\eg, regaining lost weight), they may join the app again starting a new life (Section~\ref{subsec:intent_performance}). In many cases, the primary intent of their new life remains the same as before (Section~\ref{subsec:intent_multiple_lives}).
\end{enumerate}

\subsection{Implications for Increasing Engagement}
Our findings around the novel multiple live paradigm have implications for increasing user engagement in activity tracking applications.
For example, early recognition of the primary user intent could enable a more engaging personalized user experience.
We find that it only takes 6 days from when a user starts using the app to predict a user's eventual primary activity with 80\% accuracy.
Once the user intent is inferred, one could cross-promote app features that support the user's primary intent (\eg, walking, running, or cardio for users attempting to lose weight).

Furthermore, observing how different user intents lead to user engagement with varying number of lives and lifetime can shine a light on how different features are used and where users see or do not see value. 
Consider the four quadrants of Figure~\ref{fig:most_freq_activity_vs_lifetime_vs_prob_of_coming_back}.
The bottom left quadrant, containing the primary intents drink and food, corresponds to few and short lifetimes.
This suggests that these experiences in the app are currently not well supported.
The top left quadrant, containing primarily heart rate-logging users, shows that these users tend to log heart rate across multiple lives, but only for a relatively short time each. 
These may be activities that are useful to track, but do not necessarily need to be tracked regularly over long periods of time since they do not vary as dynamically and quickly (\eg, resting and maximum heart rate changes rather slowly).
Here, one could consider interventions targeted at getting previous users with such intents to re-engage (\eg, notifications or e-mails such as ``We saw you have not checked your resting heart rate in a while. It might have changed!'').
The top right quadrant contains walking, weight, and logging of other activities which are associated with both more and longer lives. 
Likely, users find value in these features as they use them for a long, consecutive time, and often re-engage after prolonged inactivity.
User intents in the bottom right quadrant (run, sleep, and cardio) correspond to intents associated with long but few lives. 
In these cases, users find value in tracking these behaviors for a long, consecutive time, but once they have completed their life, they are less likely to re-engage again. 
Here, one can consider, for example, performance/goal-driven interventions such as ``Can you still run a 10k in 53 minutes?'' or ``Can we help you manage your weight?''.

Lastly, gamification techniques such as badges and rewards could be used to incentivize specific multiple-life behaviors. 
For example, they could be awarded for each re-engagement after prolonged inactivity. 
Importantly, this would need to be complemented with rewards for long-time use.
Otherwise, users may feel incentivized to leave the app when they would not have done so otherwise.

In summary, modeling usage periods as multiple lives can enable a better understanding of the user engagement mechanisms in mobile activity tracking applications and support of the user's motivations and intents.

\xhdr{Limitations}
This study is a large-scale case study of the mobile activity tracking application Argus.
Therefore, we are limited in our ability to generalize our findings and the multiple life paradigm to other mobile health applications. 
However, Argus is one of the most popular and general activity tracking apps on the market. 
Futher, our results show that multiple lives are likely due to real-world intentions and goals, and therefore are influenced by external factors. 
Therefore, our discoveries may generalize to other mobile health applications, where engagement is also driven by real-world intentions and factors.
In fact, we are currently working with another dataset from the mobile app MyFitnessPal by Under Armour, used for calorie counting and weight loss. 
In this dataset we also observe that user engagement is clearly segmented into multiple lives.



\subsection{Conclusion}


User engagement is important to the success of mobile health applications and has historically been modeled as a single lifetime of user engagement.
In this paper, we study user engagement patterns of over a million users over 31 months in a mobile activity tracking application.
In contrast to previous work on modeling user engagement in online contexts, 
we discover that user engagement in activity tracking applications exhibits \emph{multiple lives}.
We demonstrate that multiple lives occur because users set new goals, experience setback from previous progress, or are influenced by external factors (\eg, seasonal effects).
We discuss implications of the multiple life paradigm on increasing user engagement within activity tracking applications and propose a novel prediction task of predicting the number of lives of a user as well as their lifetime. 
We show that predictive models based on the insights developed in this work can successfully predict whether a user will re-engage with the app and the length of their lifetime.
Our work has implications for modeling user re-engagement in mobile health applications, and consequences for how notifications, recommendations as well as gamification can be used to increase engagement.




\begin{acks}
This research has been supported in part by 
NIH BD2K, DARPA NGS2, ARO MURI, IARPA HFC,
Stanford Data Science Initiative, and 
Chan Zuckerberg Biohub.
\end{acks}




\clearpage
\pagebreak
\balance


\bibliographystyle{ACM-Reference-Format}
\bibliography{refs}











\end{document}